\documentclass[twocolumn,english,aps,pra,eqsecnum]{revtex4-1}
\usepackage[T1]{fontenc}
\usepackage[latin9]{inputenc}
\usepackage{bm}
\usepackage{amsmath}
\usepackage{amssymb}
\usepackage{graphicx}
\usepackage{esint}

\makeatletter

\DeclareRobustCommand{\greektext}{%
  \fontencoding{LGR}\selectfont\def\encodingdefault{LGR}}
\DeclareRobustCommand{\textgreek}[1]{\leavevmode{\greektext #1}}
\DeclareFontEncoding{LGR}{}{}
\DeclareTextSymbol{\~}{LGR}{126}
\newcommand{\lyxmathsym}[1]{\ifmmode\begingroup\def\b@ld{bold}
  \text{\ifx\math@version\b@ld\bfseries\fi#1}\endgroup\else#1\fi}

 \@ifundefined{textcolor}{}
 {%
   \definecolor{BLACK}{gray}{0}
   \definecolor{WHITE}{gray}{1}
   \definecolor{RED}{rgb}{1,0,0}
   \definecolor{GREEN}{rgb}{0,1,0}
   \definecolor{BLUE}{rgb}{0,0,1}
   \definecolor{CYAN}{cmyk}{1,0,0,0}
   \definecolor{MAGENTA}{cmyk}{0,1,0,0}
   \definecolor{YELLOW}{cmyk}{0,0,1,0}
 }

\makeatother

\usepackage{babel}
\begin{document}

\title{Quantum dynamics in ultra-cold atomic physics}

\author{Q. Y. He, M. D. Reid, B. Opanchuk, R. Polkinghorne, Laura E. C. Rosales-Zárate,
P. D.~Drummond}

\affiliation{Centre for Atom Optics and Ultrafast Spectroscopy, Swinburne University
of Technology, Melbourne 3122, Australia}
\begin{abstract}
We review recent developments in the theory of quantum dynamics in
ultra-cold atomic physics, including exact techniques, but focusing
on methods based on phase-space mappings that are applicable when
the complexity becomes exponentially large. These phase-space representations
include the truncated Wigner, positive-P and general Gaussian operator
representations which can treat both bosons and fermions. These phase-space
methods include both traditional approaches using a phase-space of
classical dimension, and more recent methods that use a non-classical
phase-space of increased dimensionality. Examples used include quantum
EPR entanglement of a four-mode BEC, time-reversal tests of dephasing
in single-mode traps, BEC quantum collisions with up to $10^{6}$
modes and $10^{5}$ interacting particles, quantum interferometry
in a multi-mode trap with nonlinear absorption, and the theory of
quantum entropy in phase-space. We also treat the approach of variational
optimization of the sampling error, giving an elementary example of
a nonlinear oscillator.
\end{abstract}
\maketitle

\section{Introduction}

Quantum dynamics is one of the most fundamental problems in modern
physics. This is because time-evolution is the basis for any theoretical
prediction. Yet many-body complexity makes this an extremely challenging
task in quantum systems. New theoretical methods are needed, and quantitative
experiments with well-understood interactions are vitally important
in order to test predictions. In this article, we review some recent
developments relevant to ultra-cold atomic physics.

Ultra-cold atoms provide an exceptionally simple and well-understood
physical environment, allowing quantitative tests of dynamical theoretical
predictions \cite{prl-84-5029,nat-417-529}. Recent experiments explore
temperatures below $1nK$ \cite{sci-310-1513}, capable of demonstrating
dynamical behavior in many-body systems in new regimes. The important
new feature of these systems is that they allow isolated, macroscopic
quantum systems to evolve almost unitarily, with very little coupling
to external reservoirs. It is this feature of these experiments which
is highly unique, and not found in most previous condensed matter
experiments \cite{Datta-Electronic}. 

\begin{figure}
\includegraphics[width=0.8\columnwidth]{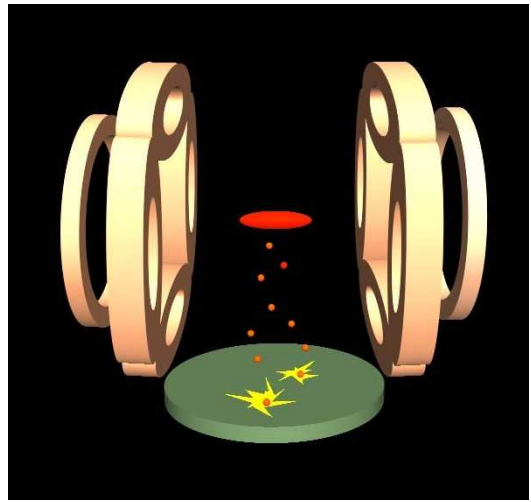}

\caption{Schematic diagram of atom counting experiments using metastable Helium
and multi-channel plate counters.\label{fig:Schematic-diagram-MCP}}
\end{figure}

Features of recent experiments include \cite{Metcalf-Laser}: 
\begin{itemize}
\item Bose-Einstein condensates: atom `photons' 
\item Atom lasers, atomic diffraction, interferometers.. 
\item Quantum superfluid fermions: atom `electrons' 
\item {Universality}: Strongly interacting fermions 
\item {Superchemistry}: Ultracold molecule formation
\item {Squeezed BEC}: Spin-squeezing with spinor atoms 
\end{itemize}
An important development is the growing ability of experimentalists
to measure atomic correlations \cite{sci-310-648} and perform atom
counting experiments with noise levels below the standard quantum
limit of Poissonian fluctuations. A typical schematic picture is shown
in Fig (\ref{fig:Schematic-diagram-MCP}), which shows a magnetically
trapped ultracold atomic cloud. After a dynamical quantum collision
of two Bose condensates, the trap is turned off and atoms are counted
by the multi-channel plate (MCP) \cite{sci-292-461} below the trap.

As well as these atomic correlation experiments, other experiments
of interest include quantum collision \cite{prl-89-020401} and quantum
interferometry experiments. In all these cases, there is an external
Hamiltonian which can be changed non-adiabatically, leading to quantum
dynamical evolution in the many-body system. This is obtained by externally
control of laser or magnetic fields.

Unlike traditional condensed matter environments, these experiments
are carried out in a high vacuum, using optical or magnetic trapping
potentials. It is this feature that allows these systems to evolve
with almost no contact with a heat reservoir. In summary, for the
first time in physics, we have large many-body quantum systems capable
of unitary evolution with a wide variety of controlled interactions.
This creates an unrivaled opportunity for testing calculations of
quantum dynamics.

\section{General Hamiltonian}

The Hamiltonian of the relevant ultra-cold atomic systems usually
are rather simple, being comprised of well-defined single-particle
and interaction terms. Thus,
\begin{equation}
{H}={H}_{0}+{H}_{I}\,
\end{equation}
 where ${H}_{0}$ and ${H}_{I}$ are the non-interacting and interacting
parts of the Hamiltonian respectively, so that $H_{0}$ is a general
linear Hamiltonian, given by: 
\begin{equation}
{H}_{0}=\sum_{ss'}\int{\Psi}_{s}^{\dagger}\left(\mathbf{r}\right)\left[V_{ss'\mathbf{r}}-\frac{\hbar^{2}\delta_{ss'}}{2m_{s}}\nabla^{2}\right]{\Psi}_{s'}\left(\mathbf{r}\right)d^{3}\mathbf{r}\,.
\end{equation}
 where ${\Psi}_{s}\left(\mathbf{r}\right)$ is a quantum field operator
$\mathbf{r}$ with internal spin or atomic species index $s=1,\ldots S$,
where $:\ldots:$ indicates normal ordering, and we use the Einstein
summation convention for repeated indices \cite{Berezin-Method}.
In addition, $m_{s}$ is the mass of species $s$, $V_{ss'\mathbf{r}}$
is a local potential, and ${H}_{I}$ describes particle-particle interactions
with interaction potential $U_{ss'\mathbf{r}\mathbf{r}'}$ : 
\begin{equation}
{H}_{I}=\frac{1}{2}\sum_{ss'}\int\int:\left|{\Psi}_{s}\left(\mathbf{r}\right)\right|^{2}U_{ss'\mathbf{r}\mathbf{r}'}\left|{\Psi}_{s'}\left(\mathbf{r}'\right)\right|^{2}:d^{3}\mathbf{r}d^{3}\mathbf{r}'\,.
\end{equation}

It is convenient to introduce local mode operators to treat such quantum
field equations\cite{drummond_carter_87}. We describe this here for
definiteness, although a more general mode expansion can be used.
We introduce $\mathcal{M}=SM^{3}$ fermionic or bosonic annihilation
operators $\tilde{\bm{a}}=\left(\tilde{a}_{\mathbf{k}s}\right)$ in
momentum space, labelled by momentum $\left(\mathbf{k}=\Delta k\mathbf{j}\right)$
and spin $\left(s\right)$. Here we assume periodic boundaries in
a finite volume $V=L^{3}$, and a lattice of $M^{3}$ cells, with
momentum spacing of $\Delta k=2\pi/L$ in each coordinate. Localized
annihilation and creation operators ${a}_{n}$ on a spatial lattice
of position indices $\mathbf{r}_{\mathbf{n}}$, with cell volume $\Delta V=V/M^{3}$,
are defined using a discrete Fourier transform: 
\begin{align}
{a}_{n} & =\frac{1}{M^{3/2}}\sum_{\mathbf{k}}\tilde{a}_{\mathbf{k}s}\exp\left[\frac{2\pi i\mathbf{k}\cdot\mathbf{n}}{\Delta kM}\right]
\end{align}
The combined index $n$ is a spin-space 4-vector, $n\equiv\left(\mathbf{n},s\right)$.
In the case of bosonic (fermionic) fields, the commutators (anticommutators)
are defined as: 
\begin{eqnarray}
\left[{a}_{m},{a}_{n}^{\dagger}\right]_{\pm} & = & \delta_{mn}^{4}\nonumber \\
\left[{a}_{m},{a}_{n}\right]_{\pm} & = & 0\,\,.
\end{eqnarray}

The corresponding local number operator is ${N}_{m}={a}_{m}^{\dagger}{a}_{m}$.
The continuum Hamiltonian is regained in the limit of a large number
of lattice sites of the resulting Hubbard type model Hamiltonian:

\begin{equation}
{H}({\bm{a}}^{\dagger},{\bm{a}})=\lim_{\Delta V\rightarrow0}\hbar\sum_{nn'}\left[\omega_{nn'}{a}_{n}^{\dagger}{a}_{n'}+\frac{1}{2}\chi_{nn'}:{N}_{n}{N}_{n'}:\right]\,.
\end{equation}
 In the uniform case, the hopping matrix $\omega_{nn'}$ is 
\begin{equation}
\omega_{nn'}=\frac{\hbar}{2m_{s}}\sum_{\mathbf{k}}\mathbf{k}^{2}\exp\left[\frac{2\pi i\mathbf{k}\cdot\left(\mathbf{n}-\mathbf{n}'\right)}{\Delta kM}\right]\delta_{ss'}
\end{equation}
and the interaction matrix $\chi_{\mathbf{mn}}$ is (approximately):
\[
\hbar\chi_{nn'}=U_{ss'\mathbf{r}_{\mathbf{n}}\mathbf{r}_{\mathbf{n}'}}\,.
\]
 While this is introduced here as an approximation to a continuum
system, it is also experimentally feasible to use an optical lattice
\cite{prl-82-2022,jpb-35-3095} to engineer the hamiltonian directly,
so that each spatial index coincides with a local trapping potential
well. In this way one can obtain a physical system directly corresponding
to the famous Hubbard model of condensed matter \cite{rsa-276-238},
except that it does not involve the numerous approximations that would
be required in a typical condensed matter setting. In the following
calculations, we will assume for simplicity that the interaction is
local in space, with $\chi_{nn'}=\chi_{ss'}\delta_{\mathbf{nn'}}^{3}$;
although this is not essential.

\subsection{Exponential complexity}

The central issue that makes quantum dynamical calculations difficult
in many body quantum physics is the issue of exponential complexity
\cite{itp-21-467}. For example, if we consider $N$ bosons distributed
among $\mathcal{M}$ modes, the number of distinct orthogonal quantum
states is obtained by combinatorics: how many ways can we divide the
particles amongst the modes? The number of quantum states is then
simply: 
\begin{equation}
N_{B}=\frac{(\mathcal{M}+N-1)!}{(\mathcal{M}-1)!N!}
\end{equation}

To give relevant numbers, suppose we consider numbers that are approximately
typical of many ultra-cold atom experiments, with $N=\mathcal{M}=500,000$.
One finds that: 
\begin{equation}
N_{B}\approx2^{2\mathcal{M}}\approx10^{300,000}
\end{equation}
 With fermions, we have fewer states, since each mode can have an
occupation number of $0$ or $ $1, meaning that: 
\begin{equation}
N_{F}\approx2^{\mathcal{M}}\approx10^{150,000}
\end{equation}
 In either case, there are more linear equations to solve than atoms
in the universe. By comparison, the number of classical equations
would be\textbf{ 
\begin{equation}
N_{C}=2\mathcal{M}=10^{6}
\end{equation}
 }

While the classical problem is difficult, it is soluble on many current
digital computers. The quantum problem, on the other hand, is nearly
impossible to treat. In particular, one can't diagonalize the Hamiltonian,
which is now a $10^{300,000}\times10^{300,000}$ matrix in the bosonic
case.

\section{Exact dynamics}

If there are small numbers of modes, one can indeed obtain exact eigenvectors.
In this case it is possible to diagonalize the Hamiltonian, and obtain
the eigenvectors and eigenvalues for a small number of particles,
typically in the range $10-100$. As an example, we consider the generation
of entanglement through four-mode nonlinear dynamics in two-well trap
holding a two-species BEC systems, in which the nonlinearity enters
through S-wave scattering interactions \cite{EPR}. The basic interaction
that generates entanglement in the first place is the nonlinear S-wave
scattering interaction, which we consider to be an interaction between
two spin-states in $^{87}Rb$. The two spin states labeled $i=1,2$
are $|1\rangle\equiv|F=1,\, m_{F}=+1\rangle,\ |2\rangle\equiv|F=2,\, m_{F}=-1\rangle$,
and there are two spatial modes corresponding to optical trapping
of modes labeled $a,b$ for clarity. Thus, $S=2$, $M=2$ and $\mathcal{M}=SM=4$.
With such a small number of modes there is no exponential complexity
issue, and the Hamiltonian can be exactly diagonalized using numerical
techniques.

The Hamiltonian for the coupled system is:
\begin{equation}
{H}/\hbar=\omega\sum_{i}{a}_{i}^{\dagger}{b}_{i}+\frac{1}{2}\left[\sum_{ij}\chi_{ij}{a}_{i}^{\dagger}{a}_{j}^{\dagger}{a}_{j}{a}_{i}\right]+\left\{ {a}_{i}\leftrightarrow{b}_{i}\right\} \,.\label{eq:Hamiltonian-1}
\end{equation}
Here $\omega$ is the inter-well tunneling rate between the two wells
with localized modes ${a}_{i},{b}_{i}$ while $\chi_{ij}$ is the
intra-well interaction matrix between the different spin components. 

We can solve this using either Schroedinger or Heisenberg equations
of motion. To illustrate this, suppose that $\omega=0$, and we have
just one well. In the Heisenberg case, one obtains:
\begin{eqnarray}
\frac{d{a}_{i}}{dt} & = & \frac{i}{\hbar}\left[{H},{a}_{i}\right]\nonumber \\
 & = & -i\sum_{j}\chi_{ij}{N}_{j}{a}_{i}.
\end{eqnarray}

Since the number of particles is conserved in each mode, this has
the solution:
\begin{eqnarray}
{a}_{i}\left(t\right) & = & \exp\left[-i\sum_{j}\chi_{ij}{N}_{j}t\right]{a}_{i}\left(0\right)
\end{eqnarray}

More generally, it is convenient to use a matrix expansion of the
Hamiltonian in a number-state basis. For dynamics, we explicitly assume
that ${a}_{1},\ b_{1}$ and $a_{2},\ b_{2}$ are initially in coherent
states. This models the relative coherence between the wells obtained
with a low inter-well potential barrier, together with an overall
Poissonian number fluctuation as typically found in an experimental
BEC. We note that the coherent state also includes an overall phase
coherence, which has no effect on our results. For simplicity, we
suppose that the initial state is prepared in an overall four-mode
coherent state using a Rabi rotation: $|\psi>=|\alpha>_{{a}_{1}}|\alpha>_{b_{1}}|\alpha>_{a_{2}}|\alpha>_{b_{2}}$.

Next, we assume that the inter-well potential is increased so that
each well evolves independently. Finally, we decrease the inter-well
potential for a short time, so that it acts as a controllable, non-adiabatic
beam-splitter \cite{olsen}, to allow interference between the wells,
followed by independent spin measurements in each well.

\subsection{\textit{\emph{Squeezed and entangled states produced by double-well
BEC}}}

We now use the techniques given above to investigate a particular
dynamical strategy for generating EPR entanglement. The technique
treated here is generally along the lines investigated experimentally
in fibre-optics, by comparison with squeezing and entanglement experiments
in optical fibers \cite{fiberexperimentNature,FiberEntangle,FiberEntangleLeuchs}.
An important difference is that the fiber experiments use time-delayed
pulses to eliminate interactions between the components. This is not
readily feasible in BEC experiments, although Feshbach resonances
can achieve this to some extent. 

Let ${a}_{1},$$a_{2}$ be operators for two internal states in the
$A$ well and $b_{1}$ , $b_{2}$ operators for two internal states
at the $B$ well. $N_{A}=a_{2}^{\dagger}a_{2}+{a}_{1}{a}_{1}$ and
$N_{B}=b_{2}^{\dagger}b_{2}+b_{1}^{\dagger}b_{1}$ are the atom number
operators of these modes in each well. We define Schwinger spin operators
at each site for the measurement of the EPR paradox and entanglement.
We define general, phase-rotated spin components according to: 
\begin{eqnarray}
J_{x}^{A} & = & \left(a_{2}^{\dagger}{a}_{1}e^{i(\theta_{2}-\theta_{1})}+{a}_{1}a_{2}e^{-i(\theta_{2}-\theta_{1})}\right)/2\ ,\nonumber \\
J_{y}^{A} & = & \left(a_{2}^{\dagger}{a}_{1}e^{i(\theta_{2}-\theta_{1})}-{a}_{1}a_{2}e^{-i(\theta_{2}-\theta_{1})}\right)/2i\ ,\nonumber \\
J_{z}^{A} & = & \left(a_{2}^{\dagger}a_{2}-{a}_{1}{a}_{1}\right)/2\ 
\end{eqnarray}
at $A$ and similar definition at $B$, while $\Delta\theta=\theta_{2}-\theta_{1}$
is the phase shift between mode $1$ and mode $2$.

\begin{figure}[h]
\begin{centering}
\includegraphics[width=0.9\columnwidth]{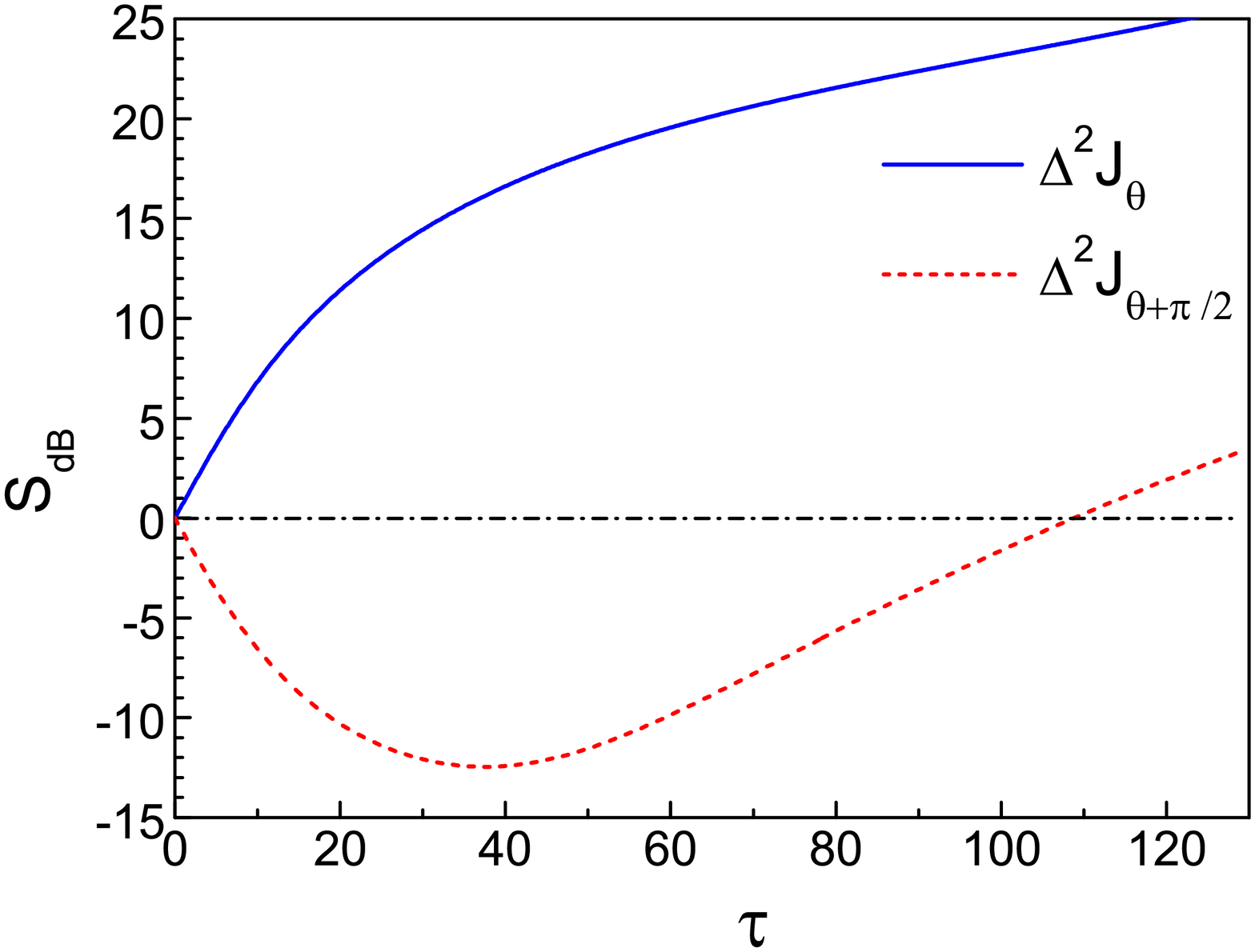}
\par\end{centering}

\centering{}\caption{Squeezing of Schwinger spin operators $10log_{10}\left(\Delta^{2}J_{\theta}/n_{0}\right)$
(solid), $10log_{10}\left(\Delta^{2}J_{\theta+\pi/2}/n_{0}\right)$
(dashed), and shot noise level $n_{0}=|\langle J_{y}\rangle|/2$ (dash-dotted)
via BEC with number of $Rb$ atoms. Here the parameters correspond
to $Rb$ atoms at magnetic field $B=9.131G$, with scattering lengths
${a}_{1}=100.4a_{0}$, $a_{22}=95.5a_{0},$ and ${a}_{1}=80.8a_{0}$.
$a_{0}=53pm$. The coupling constant $\chi_{ij}\propto2\lyxmathsym{\textgreek{w}}_{\perp}a_{ij}$.
Here $N_{A}=200$, $\tau=\chi_{11}N_{A}t$.\label{fig:squeezing-after-BEC.} }
\end{figure}

We the select the phase shift to make sure $\langle J_{y}\rangle\neq0$.
The Schwinger spin operators orthogonal to $J_{y}$ are given as $J(\theta)=cos(\theta)J_{z}+sin(\theta)J_{x}$,
all of which have the property $\langle J(\theta)\rangle=0$. We define
$\Delta\theta=\pi/2-\alpha$ where $\alpha$ is time dependent, such
that $\langle a_{2}^{\dagger}{a}_{1}\rangle=|\langle a_{2}^{\dagger}{a}_{1}\rangle|e^{i\alpha}$.
This plane contains an infinite family of maximally conjugate Schwinger
spin operators, generally given by $J(\theta)$ and $J(\theta+\pi/2)$
which obey the uncertainty relation 

\begin{equation}
\Delta^{2}J(\theta)\Delta^{2}J(\theta+\pi/2)\geq|\langle J_{y}\rangle|^{2}/4\ .
\end{equation}
Thus a state which obeys 

\begin{equation}
\Delta^{2}J(\theta)<|\langle J_{y}\rangle|/2<\Delta^{2}J(\theta+\pi/2)
\end{equation}
is a squeezed state, as shown in the Fig. \ref{fig:squeezing-after-BEC.}.
Here, we have optimized the phase choice $\theta$ to get the best
squeezing of the Schwinger spin operators by the criterion that $\partial\Delta^{2}J(\theta)/\partial\theta=0$,
hence obtaining 
\begin{equation}
tg(2\theta)=2\langle J_{z},\ J_{x}\rangle/(\Delta^{2}J_{z}-\Delta^{2}J_{x}).
\end{equation}

\begin{figure}[h]
\begin{centering}
\includegraphics[width=0.9\columnwidth]{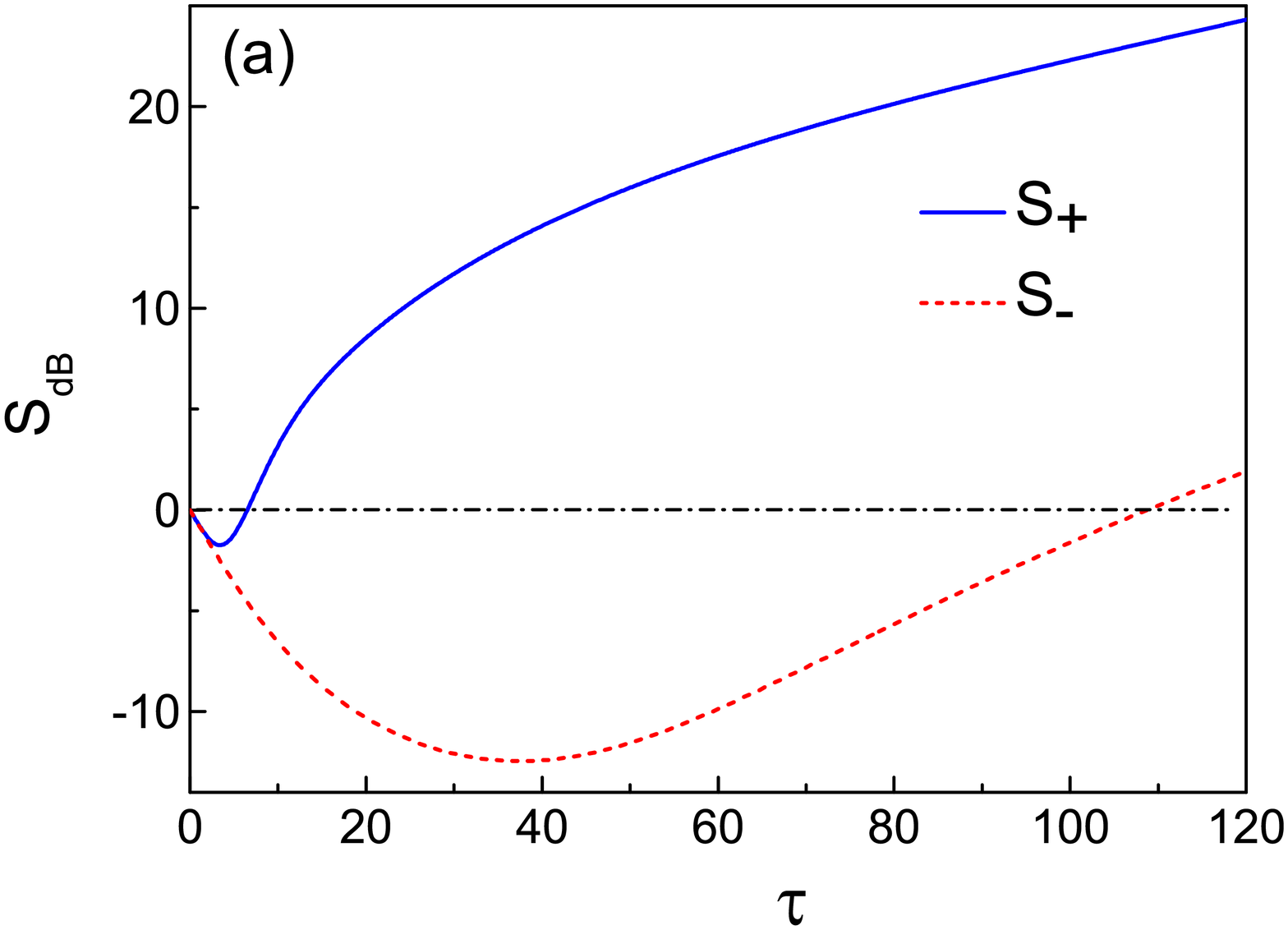}
\par\end{centering}

\begin{centering}
\includegraphics[width=0.9\columnwidth]{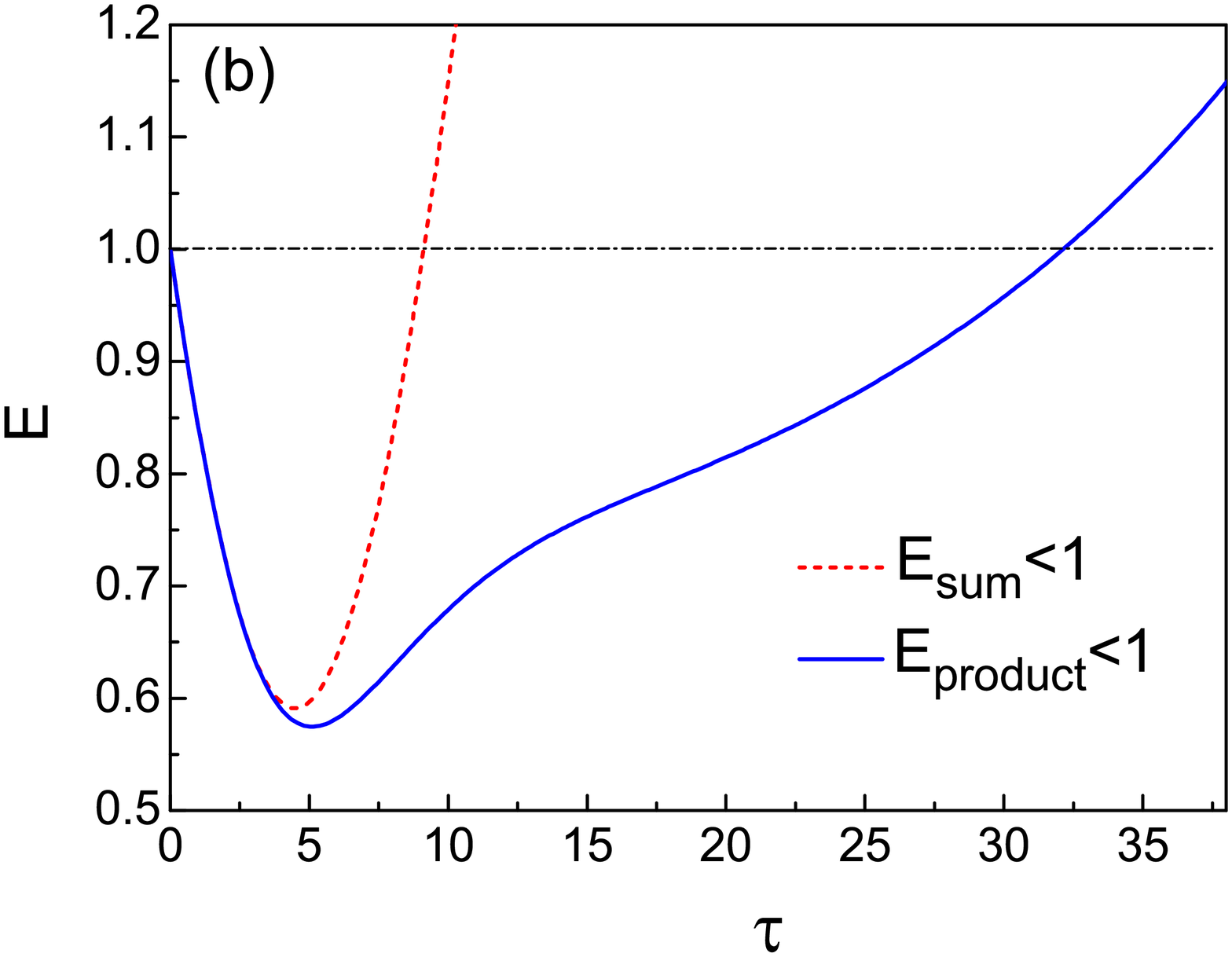}
\par\end{centering}

\centering{}\caption{(a) Squeezing of Schwinger spin operators $S_{dB}$: $S_{+}=10\, log_{10}\left[\Delta^{2}({J}_{\theta}^{A}-{J}_{\theta}^{B})/n_{0}\right]$
(solid), $S_{-}=10\, log_{10}\left[\Delta^{2}({J}_{\theta+\pi/2}^{A}+{J}_{\theta+\pi/2}^{B})/n_{0}\right]$
(dashed), and $n_{0}=(|\langle{J}_{y}^{A}\rangle|+|\langle{J}_{y}^{B}\rangle|)/2$
is shot noise (dash-dotted). (b) Entanglement ($E_{product}$) based
on the criterion (\ref{eq:product form}) by the solid curve and $E_{sum}$
in sum criterion (\ref{eq:SUM}) by the dashed curve. (This figure
has been published in Ref. \cite{EPR})\label{fig:Inference-squeezing-()}}
\end{figure}

In this strategy, spin-squeezing at each site can be readily obtained,
by unitary evolution from an initial coherent state, under the local
Hamiltonian. Then we can obtain sum and difference spins between two
sites: $\Delta^{2}{J}_{\theta\pm}^{AB}=\Delta^{2}({J}_{\theta}^{A}-{J}_{\theta}^{B})$
and $\Delta^{2}{J}_{(\theta+\pi/2)\pm}^{AB}=\Delta^{2}({J}_{\theta+\pi/2}^{A}+{J}_{\theta+\pi/2}^{B})$,
prior to using the beam-splitter - which is achieved by a modulation
of the inter-well potential, as shown in Fig. \ref{fig:Inference-squeezing-()}(a). 

After using the beam-splitter, entanglement can be detected via spin
measurements using the spin version of the Heisenberg-product entanglement
criterion \cite{FiberEntangle} 
\begin{equation}
E_{product}=\frac{2\sqrt{\Delta^{2}{J}_{\theta\pm}^{AB}\cdot\Delta^{2}{J}_{(\theta+\pi/2)\pm}^{AB}}}{|\langle{J}_{y}^{A}\rangle|+|\langle{J}_{y}^{B}\rangle|}<1\,,\label{eq:product form}
\end{equation}
or the sum criterion \cite{BowenPolarizationEnt}
\begin{equation}
E_{sum}=\frac{\Delta^{2}{J}_{\theta\pm}^{AB}+\Delta^{2}{J}_{(\theta+\pi/2)\pm}^{AB}}{|\langle{J}_{y}^{A}\rangle|+|\langle{J}_{y}^{B}\rangle|}<1\,.\label{eq:SUM}
\end{equation}
as shown in Fig. \ref{fig:Inference-squeezing-()}(b). 

While diagonalization is possible for even larger particle numbers,
the two-mode approximation becomes less and less reliable. For larger
numbers of particles the interaction energy becomes as large as as
the harmonic oscillator mode spacing. This means that few mode approximations
become inapplicable, and the problem quickly develops exponentially
large numbers of many-body states. Methods for treating this more
general case are given in the next section.

\section{Classical phase-space}

Early techniques for calculating the dynamics and thermal equilibrium
states of large quantum systems used techniques based on mappings
to a classical phase-space \cite{Gardiner-Quantum}. These were used
not just in statistical many-body theory, but also in applications
involving coherence theory and lasers. The most widely used methods
of this type are the Wigner representation\cite{wigner_32} and the
Glauber-Sudarshan \cite{glauber_63,sudarshan_63} P-representations.
These differ in some technical details. In essence, both can be used
for mapping second-quantized bosonic fields to a classical phase-space.
However, the Wigner representation corresponds to symmetrically-ordered
operator mappings, while the P-representation corresponds to normal
ordering. It is also possible to use an anti-normal ordered mapping,
called the Husimi Q-function \cite{husimi_40}, but this is less commonly
used for dynamical calculations. When used for quantum fields, in
a truncation approximation described below, these types of phase-space
method are sometimes called c-field techniques. Phase-space methods
for quantum fields were used to predict quantum squeezing in solitons
in fiber optics \cite{carter_drummond_87,carter_drummond_91,drummond_hardman_93,drummond_shelby_93,Drummond2001a},
with an excellent agreement with subsequent experimental tests\cite{Rosenbluh1991a,Heersink2005a,corney_drummond_06a}.
They were later applied to ultra-cold atomic BEC dynamics \cite{steel_olsen_98},
and have been widely used, especially at finite temperatures \cite{blakie_bradley_08,Egorov2011}.

\subsection{Glauber-Sudarshan P-representation}

This approach uses an overcomplete set of coherent states \cite{prx-131-2766,pra-59-1538},
parameterized by a complex vector $\bm{\alpha}$: 
\begin{equation}
\left|\bm{\alpha}\right\rangle =\exp\left[\bm{a}^{\dagger}\cdot\bm{\alpha}-\bm{\alpha}^{*}\cdot\bm{\alpha}/2\right]\left|\bm{0}\right\rangle 
\end{equation}
 and one then obtains an expansion for the density matrix in the form:
\begin{equation}
{\rho}=\int P(\bm{\alpha}){\Lambda}_{1}\left(\bm{\alpha}\right)d^{2M}\bm{\alpha}\label{eq:P-rep}
\end{equation}
 where ${\Lambda}_{1}\left(\bm{\alpha}\right)$ is a diagonal coherent
state projection operator, which is the basis for the expansion, defined
as: 
\begin{equation}
{\Lambda}_{1}\left(\bm{\alpha}\right)=\left|\left(-\right)\bm{\alpha}\right\rangle \left\langle \bm{\alpha}\right|\label{eq:P-basis}
\end{equation}

Here we note that the same expansion can be used for either fermions
of bosons. In the case of fermions, $\bm{\alpha}$ is a Grassmann
variable\cite{Berezin-Method}, and one must use the bracketed minus
sign in Eq (\ref{eq:P-basis}). This representation generates normal-ordered
operator products, in the sense that moments of $P(\bm{\alpha})$
correspond directly to expectation values of normally ordered operator
products.

The advantage of this approach is that it maps quantum states into
\emph{$\mathcal{M}$} complex coordinates,\textbf{ $\bm{\alpha}=\mathbf{p}+i\mathbf{x},$}
and hence only has classical complexity. Another advantage is that
the use of normally-ordered products means that there is no UV vacuum
divergence in the expectation values. However, for many quantum states,
including all entangled states, the distribution is not positive,
and indeed is highly singular.

In the bosonic case, the operator basis can be written in an alternative
form \cite{pra-68-63822}, as: 
\begin{equation}
{\Lambda}_{s}(\bm{\lambda})=\left[\frac{2}{1+s}\right]^{\mathcal{M}}:\exp\left[-2\delta{\bm{a}}^{\dagger}\delta{\bm{a}}/\left(1+s\right)\right]:\,,
\end{equation}
 where $\delta{\bm{a}}={\bm{a}}-\bm{\alpha}$ , $\delta{\bm{a}}^{\dagger}={\bm{a}}^{\dagger}-\bm{\alpha}^{*}$
are relative displacements, and $s$ indicates the operator ordering
\cite{Cahill1969}. Here $s=1$ is used for normal ordering, as in
the case of the P-representation, and other orderings are treated
in the next subsection. This allows us to recognize that the basis
is just a Gaussian function of the mode operators: an exponential
of a quadratic function of annihilation and creation operators. Just
as with similar Gaussian bases for ordinary complex functions, such
an operator basis has more than one possible form, obtained by changing
the variance.

\subsection{Wigner-Moyal phase-space}

An even older phase-space method was developed by Wigner \cite{wigner_32},
who treated thermal equilibrium problems, and Moyal \cite{moyal_49},
who extended this to a full dynamical equivalence with quantum mechanics.
This can also be written as an expansion over a Gaussian operator
basis with symmetric ordering mappings, so that $O=0$ . This reduces
the basis variance, and therefore increases the variance of the distribution
function. Formally, the expansion is written as:

\begin{equation}
\,{\rho}=\int W(\bm{\alpha}){\Lambda}_{0}\left(\bm{\alpha}\right)d^{2\mathcal{M}}\bm{\alpha}
\end{equation}

This type of distribution generates symmetrically ordered operator
products. It maps quantum states into \textbf{$\mathcal{M}$} complex
or $2\mathcal{M}$ real coordinates, and has the advantage that this
mapping gives dynamical equations that are the most similar to classical
behavior. Historically, Moyal first showed how to map quantum operators
into differential equations, and had a famous correspondence with
Dirac, who objected to the fact that the distribution had no probabilistic
interpretation.

As result, there is a problem for computational implementation. One
would like to sample the Wigner distribution probabilistically, but
this is not always possible. Since the mapping is nonpositive, there
is no generally efficient and accurate sampling procedure. The representation
is also typically UV divergent in three dimensions, due to vacuum
field fluctuations of symmetrically ordered moments.

For the case of an initial coherent state, 
\begin{equation}
\left|\Psi_{0}\right\rangle =\left|\bm{\alpha}_{0}\right\rangle 
\end{equation}
 the initial Wigner distribution is Gaussian and positive, so that
the quantum noise can be readily sampled stochastically, with: 
\begin{equation}
\bm{\alpha}=\bm{\alpha}_{0}+\delta\bm{\alpha}_{0}
\end{equation}
 where $\delta\bm{\alpha}_{0}$ is a Gaussian random complex number,
such that the only nonvanishing correlations are: 
\begin{equation}
\left\langle \delta\bm{\alpha}_{0}\delta\bm{\alpha}_{0}^{*}\right\rangle =\frac{1}{2}\,.
\end{equation}

For more general states - even as simple as a number state - the Wigner
distribution exists but has no stochastic equivalent. The quantum
dynamical manifestation of this problem is that one obtains a third-order
Fokker-Planck equation for the Wigner time evolution when there are
nonlinear Hamiltonian terms. Such an equation has no stochastic equivalent,
unless truncated to give a second order differential equation. In
this approximation, the resulting equation has a semi-classical form,
with: 
\begin{equation}
i\frac{d\alpha_{m}}{dt}=\sum_{n}\left[\omega_{mn}\alpha_{n}+\chi_{mn}\left|\alpha_{n}\right|^{2}\alpha_{m}\right]\label{eq:lattice-Wigner}
\end{equation}
 While apparently classical, this equation includes quantum noise
in the initial conditions, and can simulate nonclassical entangled
states, in an approximation which is valid in the limit of large mode
occupations

It was the nonpositivity of the Wigner representation that led Feynman
to make his famous conjecture\cite{itp-21-467} that it was not possible
to use a classical or digital computer to make a probabilistic representation
of a quantum system. Nevertheless, while not exact, the truncated
Wigner approach is relatively simple, and has useful applications
as a practical approximation when mode occupation numbers are large. 

We note that the Husimi Q-function \cite{husimi_40}, which corresponds
to antinormal ordering with $O\rightarrow-1$, is always positive.
Yet, paradoxically, this has no direct stochastic equivalent either,
since the corresponding Fokker-Planck equation is not positive-definite.
As we show in the next section, despite Feynman's remark, there are
routes to positive representations of quantum systems that \textbf{do}
have stochastic equivalents, but they involve enlarged, non-classical
phase-space mappings.

\subsection{Large-scale two-component atom interferometry}

As an illustration of the Wigner phase-space techniques, consider
interferometry of a two-component $^{87}$Rb BEC in a harmonic trap,
which is performed by many experimental groups worldwide. Atom numbers
in these experiments range from $10^{4}$ to $10^{6}$, which makes
it impossible to simulate the behavior of the system exactly in a
few mode approximation: these larger traps are inherently multi-mode.
A common approach involves the propagation of semi-classical Gross-Pitaevskii
equations \cite{Gardiner-Quantum}. Although these equations provide
a good description of the condensate evolution, they do not account
for quantum effects in the cloud, and cannot predict variances of
the observables. A truncated Wigner phase-space approach can be used
to obtain more accurate predictions \cite{europhys-lett-21-279,pra-58-4824,jpb-35-3599}. 

The effective Hamiltonian in three dimensions is

\begin{equation}
{H}/\hbar=\sum_{ss'}\int d^{3}\mathbf{\mathbf{x}}\left\{ {\Psi}_{s}^{\dagger}K_{ss'}{\Psi}_{s'}+\frac{\chi_{ss'}}{2}{\Psi}_{s}^{\dagger}{\Psi}_{s'}^{\dagger}{\Psi}_{s'}{\Psi}_{s}\right\} ,
\end{equation}
where $ $${\Psi}_{s}$ is an annihilation operator for spin $s$,
with the space position omitted for brevity, and $\chi_{ss'}$ is
the spin-dependent contact interaction strength. The operator $K_{ss'}$
is the single-particle Hamiltonian:

\begin{equation}
K_{ss'}=\left(-\frac{\hbar}{2m}\nabla^{2}+\omega_{s}+V_{s}(\mathbf{x})/\hbar\right)\delta_{ss'}+\tilde{\Omega}_{ss'}(t),
\end{equation}
where $m$ is the atomic mass, $V_{s}$ is the external trapping potential
for spin $s$, $\omega_{s}$ is the internal energy of spin $s$,
$\tilde{\Omega}_{ss'}$ is the time-dependent coupling term. Losses,
which can play a significant part in the evolution, can be included
into the model by adding a loss term to the master equation \cite{prl-89-140402}:

\begin{equation}
\frac{d{\rho}}{dt}=-\frac{i}{\hbar}\left[{H},{\rho}\right]+\sum_{n,\mathbf{l}}\kappa_{\mathbf{l}}^{(p)}\int d^{D}\mathbf{x}\mathcal{L}_{\mathbf{l}}^{(p)}\left[{\rho}\right],
\end{equation}
where $p$ is the number of the interacting particles in the loss
process, the vector $\mathbf{l}$ specifies the number of particles
with each spin participating in the interaction, and $\mathcal{L}$
is the operator functional:

\begin{equation}
\mathcal{L}_{\mathbf{l}}^{(p)}\left[{\rho}\right]=2{O}_{\mathbf{l}}^{(p)}{\rho}{O}_{\mathbf{l}}^{(p)\dagger}-{O}_{\mathbf{l}}^{(p)\dagger}{O}_{\mathbf{l}}^{(p)}{\rho}-{\rho}{O}_{\mathbf{l}}^{(p)\dagger}{O}_{\mathbf{l}}^{(p)}.
\end{equation}
The reservoir coupling operators ${O}_{\mathbf{l}}^{(p)}$ are the
distinct $p$-fold products of local field annihilation operators,

\begin{equation}
{O}_{\mathbf{l}}^{(p)}={O}_{\mathbf{l}}^{(p)}({\mathbf{\Psi}})={\Psi}_{l_{1}}(\mathbf{x}){\Psi}_{l_{2}}(\mathbf{x})\ldots{\Psi}_{l_{p}}(\mathbf{x}),
\end{equation}
describing local collisional losses. 

After a transformation to the Wigner representation using functional
Wigner correspondences \cite{pra-58-4824} and truncation of high-order
terms, the resulting equations take a form similar to that of coupled
GPEs with the addition of stochastic terms \cite{Egorov2011}:

\begin{align}
\frac{d\phi_{s}(\mathbf{x})}{dt} & =-i\sum_{u}\left(K_{su}\phi_{u}+U_{su}|\phi_{u}|^{2}\phi_{s}\right)\nonumber \\
 & -\Gamma_{s}(\mathbf{x})+\sum_{p,\mathbf{l}}\beta_{\mathbf{l},s}^{(p)}\zeta_{\mathbf{l}}^{(p)}(\mathbf{x},t),
\end{align}
 where $\phi_{s}(\mathbf{x})$ is a Wigner c-field corresponding to
the operator field $\Psi_{s}(\mathbf{x})$. Additionally, $\Gamma_{s}$
is the nonlinear loss term and $\beta_{\mathbf{l},s}^{(p)}$ is the
damping noise coefficient which are both functions of the Wigner fields,
while $\zeta_{\mathbf{l}}^{(p)}(\mathbf{x},t)$ is a corresponding
complex, stochastic delta-correlated Gaussian noise$ $ such that:

\begin{equation}
\left\langle \zeta_{\mathbf{l}}^{(p)}(\mathbf{x},t)\zeta_{\mathbf{k}}^{(p')*}(\mathbf{x}^{\prime},t^{\prime})\right\rangle =\delta_{\mathbf{l}\mathbf{k}}\delta_{pp'}\delta^{D}\left(\mathbf{x}-\mathbf{x}^{\prime}\right)\delta\left(t-t^{\prime}\right).
\end{equation}

These stochastic equations can be solved numerically using conventional
methods on a discrete lattice. The resulting equations then have a
similar form to the general lattice Wigner equations, Eq (\ref{eq:lattice-Wigner}),
apart from additional loss and noise terms:
\begin{align}
\frac{d\alpha_{m}}{dt} & =-i\sum_{n}\left[\omega_{mn}\alpha_{n}+\chi_{mn}\left|\alpha_{n}\right|^{2}\alpha_{m}\right]\nonumber \\
 & -\Gamma_{m}+\sum_{p,\mathbf{l}}\beta_{\mathbf{l},s}^{(p)}\zeta_{\mathbf{l},m}^{(p)}(t),
\end{align}
 Correlations of any order can be extracted, and specifically, one
can obtain the value of squeezing parameter $\xi^{2}$. This serves
as an indicator of entanglement in the condensate \cite{nat-409-63,pra-50-67}:

\begin{equation}
\xi^{2}=\frac{N\Delta{S}_{min}^{2}}{\langle{S}\rangle^{2}},
\end{equation}
where $N$ is the number of atoms, ${S}$ is the total spin, and $\Delta{S}_{min}^{2}$
is the minimal variance of spin over all possible directions. The
squeezing parameter is a fourth-order field correlation, with values
$\xi^{2}<1$ indicating entangled, or spin squeezed states. By simulating
the evolution of the condensate under different conditions one can
find the optimal regime for producing maximum squeezing. 

\begin{figure}
\includegraphics[width=0.9\columnwidth]{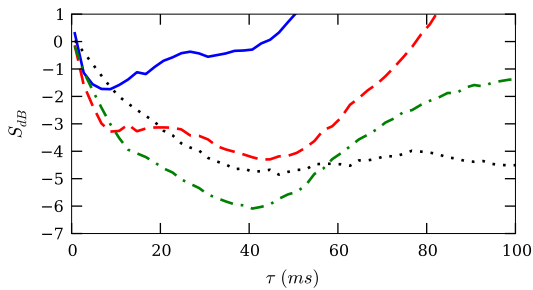}\caption{Wigner simulations of squeezing in the vicinity of $9.1\,\mathrm{G}$
Feshbach resonance in $^{87}$Rb plotted with a logarithmic scale
$S_{dB}=10\, log_{10}(\xi^{2})$. Inter-component scattering lengths:
${a}_{1}=80.0\, a_{0}$ (blue solid line), ${a}_{1}=85.0\, a_{0}$
(red dashed line), ${a}_{1}=90.0\, a_{0}$ (green dash-dotted line)
and ${a}_{1}=95.0\, a_{0}$ (black dotted line).\label{fig:ramsey-squeezing}}
\end{figure}

For example, the inter-component scattering length of optically trapped
two-component $^{87}$Rb condensate of states $|F=1,\, m_{F}=+1\rangle$
and $|F=2,\, m_{F}=-1\rangle$ near a Feshbach resonance at $9.1\,\mathrm{\mathrm{G}}$
can be changed by varying the strength of magnetic field. When one
moves closer to the resonance, the inter-component scattering length
decreases, providing better squeezing, but inter-component losses
increase correspondingly \cite{pra-80-050701}, eliminating the coherence
faster. Fig (\ref{fig:ramsey-squeezing}) shows the evolution of squeezing
parameter in time for four different values of ${a}_{1}$, where best
results are achieved for ${a}_{1}=90\, a_{0}$. In other words, one
has to pick an optimal but non-zero detuning from the Feshbach resonance
in order to get maximum squeezing.

\section{Non-classical phase-space}

We generically regard any expansion of the density matrix in a complete,
non-orthogonal basis set with continuous variables as a phase-space
method. Such methods are not restricted to classical mappings, however.
In the following sections, we review progress in developing non-classical
phase-space mappings, typically using higher than classical dimensionality.
This approach is essentially a middle ground between the low complexity
of classical phase-space, and the exponentially large complexity of
the full many-body Hilbert space.

\subsection{Positive-P function methods}

In this approach, one extends the mapping of a bosonic field theory
onto classical phase-space used in the Glauber-Sudarshan P-function,
to a larger phase-space of double the classical dimension. This is
best thought of as a minimal prescription for including coherent state
superpositions and entanglement into the basis set. Thus, one defines:

\begin{equation}
{\rho}=\int P_{+}(\bm{\alpha},\bm{\beta}){\Lambda}_{+}\left(\bm{\alpha},\bm{\beta}\right)d^{2M}\bm{\alpha}d^{2M}\bm{\beta}\label{eq:P-rep-1}
\end{equation}
 where the basis set is now: 
\begin{equation}
{\Lambda}_{+}\left(\bm{\alpha},\bm{\beta}\right)=\frac{\left|\bm{\alpha}\right\rangle \left\langle \bm{\beta}^{*}\right|}{\left\langle \bm{\beta}^{*}\right|\left|\bm{\alpha}\right\rangle }
\end{equation}

This enlarged phase-space allows positive probabilities for any quantum
state, since it is possible to prove an existence theorem that any
physical density matrix has a positive distribution in this form.
While this itself is no different to the properties of the Husimi
Q-function - another positive representation - there are additional
advantages, as we explain below. In comparison to the usual diagonal
Glauber-Sudarshan case, we note that the +P representation has these
differences: 
\begin{itemize}
\item It now maps quantum states into 4M real coordinates:\textbf{{{}
$\bm{\alpha},\bm{\beta}=\mathbf{p}+i\mathbf{x},\,\,\mathbf{p}'+i\mathbf{x}'$}} 
\item The corresponding phase-space has double the dimensionality of a classical
phase-space 
\item The advantage is that one can represent superpositions including entangled
states without singularities. 
\end{itemize}

\subsection{+P Existence Theorem }

The most significant property of the +P method is the existence theorem:
a positive P-function always exists, for any density matrix. While
the proof is too lengthy to be given here, we quote the result, which
has a simple constructive form \cite{drummond_gardiner_1980}. For
any hermitian, positive-definite density matrix ${\rho}$, there is
a corresponding positive-P distribution, of `canonical' form: 
\begin{equation}
P_{+}(\boldsymbol{\alpha},\boldsymbol{\beta})=\left[\frac{1}{4\pi^{2}}\right]^{\mathcal{M}}e^{-\left|\boldsymbol{\alpha}-\boldsymbol{\beta}^{*}\right|^{2}/4}\left\langle \frac{\boldsymbol{\alpha}+\boldsymbol{\beta}^{*}}{2}\right|{\rho}\left|\frac{\boldsymbol{\alpha}+\boldsymbol{\beta}^{*}}{2}\right\rangle 
\end{equation}

The advantage here is not just that the distribution is nonsingular,
but more importantly that probabilistic sampling is possible. This
is a crucial issue when dealing with the complexity of many-body systems.
Generally, random, probabilistic sampling is the only practical approach
for subduing the unruly nature of an exponentially complex Hilbert
space.

This approach nevertheless is not without its own problems. The use
of a non-orthogonal basis means that the distribution is non-unique.
We can choose a compact form - like the `canonical' positive form
given above - as an initial condition. However, this form is generally
not conserved by the equations of motion, since it is not unique.
If the equations of motion generate distributions that are less compact
as time evolves, the this allows sampling errors to grow with time.

An important application of the +P distribution is the calculation
of measurable operator moments. In order to calculate an operator
expectation value, there is a correspondence between the moments of
the +P distribution, and the normally ordered operator products. These
come directly from the fact that coherent state are eigenstates of
the annihilation operator, and that $\mathrm{{Tr}}\left[{\Lambda_{+}}(\boldsymbol{\alpha},\boldsymbol{\beta})\right]=1$,
which means that any normally ordered operator product is simply a
stochastic average of the phase-space variables: 
\begin{align}
\langle{a}_{m}^{\dagger}\cdots{a}_{n}\rangle & =\mathrm{{Tr}}\left[\cdots{a}_{n}{\Lambda_{+}}(\boldsymbol{\alpha},\boldsymbol{\beta}){a}_{m}^{\dagger},\ldots\right]\\
 & =\int\int P_{+}(\boldsymbol{\alpha},\boldsymbol{\beta})[{\beta}_{m}\cdots{\alpha}_{n}]d^{2\mathcal{M}}\boldsymbol{\alpha}\, d^{2\mathcal{M}}\boldsymbol{\beta}\nonumber 
\end{align}

\subsection{+P time-evolution}

The route to obtaining time-evolution equations is to map operator
equations into differential equations for the P-function. Differentiating
the +P projection operator gives the following four identities:

\begin{eqnarray}
{a}_{m}^{\dagger}{\Lambda} & = & \left[\frac{\partial}{\partial\alpha_{m}}+\beta_{m}\right]{\Lambda}\nonumber \\
{a}_{m}{\Lambda} & = & \alpha_{m}{\Lambda}\nonumber \\
{\Lambda}{a}_{m} & = & \left[\frac{\partial}{\partial\beta_{m}}+\alpha_{m}\right]{\Lambda}\nonumber \\
{\Lambda}{a}_{m}^{\dagger} & = & \beta_{m}{\Lambda}
\end{eqnarray}

Since the projector is an analytic function of both $\alpha^{m}$
and $\beta^{m}$, we can obtain alternate identities by replacing
$\partial/\partial\alpha$ by either $\partial/\partial\alpha_{x}$
or $\partial/i\partial\alpha_{y}$. This equivalence allows a positive-definite
diffusion to be obtained, with stochastic evolution. The result of
this procedure is that our exponential complex quantum problem\textbf{
}is now transformed into a stochastic equation. Thus, for the case
of a single-component Bose gas with S-wave interactions, one obtains
the following equations in the simplest case:

\begin{eqnarray}
i\frac{d\alpha^{m}}{dt} & = & \omega_{mn}\alpha_{n}+\left[\chi\alpha_{m}\beta_{m}+\sqrt{i\chi}\ \xi_{m}^{(1)}\left(t\right)\right]\alpha_{m}\\
-i\frac{d\beta_{m}}{dt} & = & \omega_{mn}\beta_{n}+\left[\chi\alpha_{m}\beta_{m}+\sqrt{-i\chi}\ \xi_{m}^{(2)}\left(t\right)\right]\beta_{m}\nonumber 
\end{eqnarray}

Here $ $$\xi_{m}^{(i)}\left(t\right)$ are $ $2M independent real,
random Gaussian noise terms, with correlations given on a discrete
spatial lattice with cell volume $\Delta V$, by:
\begin{align}
\left\langle \xi_{m}^{(i)}\left(t\right)\xi_{m'}^{(i')}\left(t'\right)\right\rangle  & =\frac{1}{\Delta V}\delta_{mm'}\delta_{ii'}\delta\left(t-t^{\prime}\right).
\end{align}
Similar techniques can also be used for the case of fermions, using
the Gaussian representation method, with some modifications which
are not treated in detail here. In both cases, the essential trade-off
is that, as with any sampling technique, many parallel trajectories
are needed to control growing sampling errors. This can be modified
by changing the choice of basis, and the stochastic mapping which
is not unique. This approach leads to the idea of a `stochastic gauge'\cite{job-5-S281},
which multiplies the basis operator ${\Lambda_{+}}$ by a random weight
$\Omega$, and improves convergence properties by reducing the sampling
error. 

We emphasize that while any stochastic method only converges for a
large number of samples, the sampling error is generally a well-controlled
numerical error. Like the momentum cut-off, the sample-size can be
changed and the error monitored using well-defined numerical procedures.

\subsection{Time-reversal tests}

We now consider two examples of the application of the +P distribution
to quantitative simulations of time-evolution under our Hubbard-type
Hamiltonian.

By choosing a modified stochastic gauge, it is possible to simulate
a very large number of bosons, and verify the simulation by carrying
out a time-reversal test. This takes advantage of the fact that unitary
evolution is time-reversible, so that changing the Hamiltonian sign
will cause a reverse evolution to occur, which should recreate the
initial state. Such tests have been carried out with up to $10^{23}$
interacting bosons\cite{dowling_drummond_05}. In Fig (\ref{fig:Time-reversal-test-quad}),
we show a time-reversal test carried out for the single mode anharmonic
oscillator with an initial coherent state of $\alpha=10$, and a corresponding
mean initial population of $N=100$ bosons. The quantity graphed is
the mean quadrature variable, defined as ${X}=\left({a}+{a}^{\dagger}\right)/2$.
The Hamiltonian sign was changed at $\tau=0.5$ , resulting in a restoration
of the initial coherence.

\begin{center}
\begin{figure}
\begin{centering}
\includegraphics[width=0.9\columnwidth]{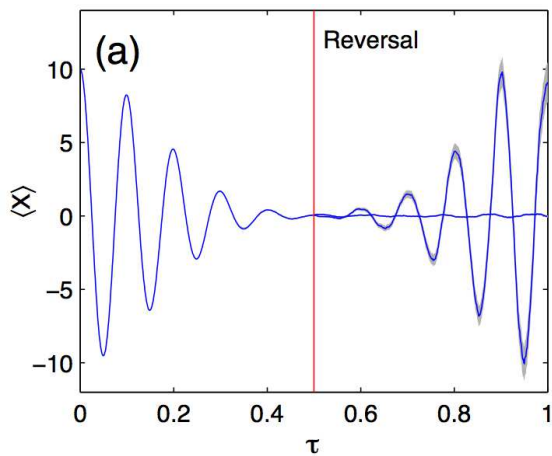} 
\par\end{centering}

\caption{Time-reversal test of anharmonic oscillator with $N=100$ initial
bosons in a coherent state. The shaded areas for $\tau>0.5$ indicate
a slowly growing sampling error-bar, calculated using the central
limit theorem and the sampled variance. \label{fig:Time-reversal-test-quad}}
\end{figure}

\par\end{center}

\begin{center}
~ 
\par\end{center}

This calculation also demonstrates a subtle feature of the +P stochastic
method, which is that the phase-space distribution is not unique!
In fact, a moment's thought serves to illustrate that this is a necessary
feature of a stochastic method that represents unitary evolution in
quantum mechanics. If the mapping has stochastic behavior during the
time-evolution, then it will spread in the phase-space as time evolves.
Technically, the phase-space entropy must increase.

\begin{figure}
\begin{centering}
\includegraphics[width=0.9\columnwidth]{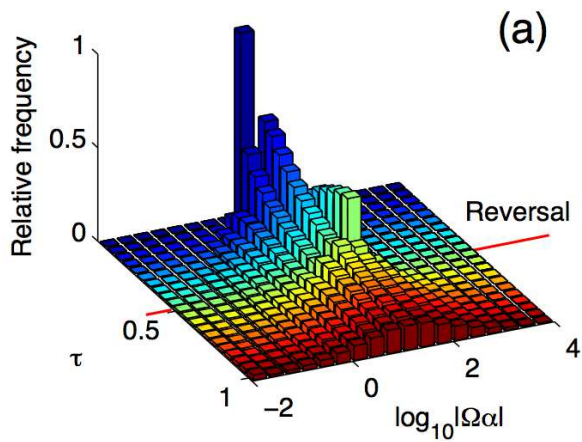} 
\par\end{centering}

\caption{Phase-space distributions, showing non-uniqueness after time-reversal
at normalized time $\tau=0.5$. The horizontal axis is a product of
the phase-space coordinate $\alpha$ and the random stochastic weight
$\Omega$. \label{fig:Phase-space-distributions}}
\end{figure}

Yet reversing time simply results in another type of unitary evolution,
also with a stochastic equivalent. The result is that the phase-space
distribution spreads even more, increasing the phase-space entropy
yet further. This is illustrated in Fig (\ref{fig:Phase-space-distributions}),
which graphs the distribution underlying the mean quadratures depicted
in Fig (\ref{fig:Time-reversal-test-quad}). Clearly, the final distribution
after time-reversal is totally different from the initial distribution,
which is a delta-function in phase-space. The final distribution after
time-reversal is a Gaussian convolution of the original delta-function.
Despite this, the two distributions are physically identical, and
have identical observable moments due to the non-uniqueness of the
basis set.

\begin{center}
~ 
\par\end{center}

\subsection{BEC collision: $10^{5}$ bosons, $10^{6}$ spatial modes}

Next we consider a case of extremely large complexity: a collision
of two Bose condensates, each with a very large number of particles
and modes.

\begin{figure}
\includegraphics[width=0.9\columnwidth]{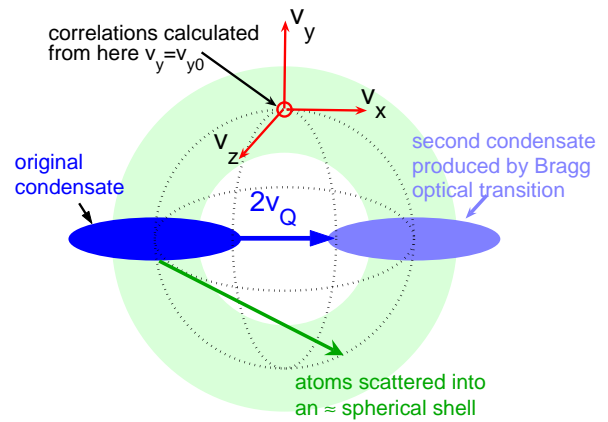}

\caption{Schematic diagram of a BEC collision. A second condensate is produced
in situ by a Bragg scattering pulse, with a relative velocity of $2v_{Q}$,
causing a quantum collision. This results in quantum correlated atoms
scattering into a spherical shell of velocities around the original
mean velocity. \label{fig:Schematic-diagram}}
\end{figure}

Recent experiments on ultra-cold Bose-Einstein condensates have been
able to generate collisions of quantum condensates with large numbers
($>10^{5}$) of particles\cite{prl-89-020401}. These experiments
typically use metastable $^{4}He^{*}$ condensates so that he particle
correlations can be readily measured. The initial state is simply
a trapped BEC in which half the atoms have been accelerated to a high
relative velocity compared to the other half using optical Bragg scattering
techniques, as shown in Fig (\ref{fig:Schematic-diagram}). Atoms
collide to produce a scattered halo of correlated atoms, involving
both spontaneous and stimulated emission into the scattered modes.
These quantitative experiments provide a rigorous test of the methodology
of these simulations.

We consider the collision\cite{deuar_drummond_07} of two pure $^{23}$Na
BECs, with a similar design to a recent experiment at MIT\cite{vogels_xu_02},
and more recent experiments in France\cite{krachmalnicoff_jaskula_10}.
A $1.5\times10^{5}$ atom condensate is prepared in a cigar-shaped
magnetic trap with frequencies 20 Hz axially and 80 Hz radially. A
brief Bragg laser pulse is used to coherently impart a velocity of
${2{\rm v_{Q}}}=19.64$ mm/s to half of the atoms, that is much greater
than the sound velocity of $3.1$ mm/s. At this point the trap is
turned off so that the wavepackets collide freely.

The coupling constant $g$ depends on the s-wave scattering length
$a$ ($2.75$nm in the case of $^{23}$Na). We begin the simulation
in the center-of-mass frame at the moment the lasers and trap are
turned off \mbox{($t=0$).} The initial wavefunction is modeled
as the coherent-state mean-field Gross-Pitaevskii (GP) solution of
the trapped $t<0$ condensate, but modulated with a factor $\left[e^{ik_{{\rm Q}}{\rm x}}+e^{-ik_{{\rm Q}}{\rm x}}\right]/\sqrt{2}$
which imparts initial velocities \mbox{${\rm v_{x}}={\rm \pm v_{Q}}=\pm\hbar k_{{\rm Q}}/m$}
in the ${\rm x}$ direction.

\begin{figure}
\includegraphics[width=1\columnwidth]{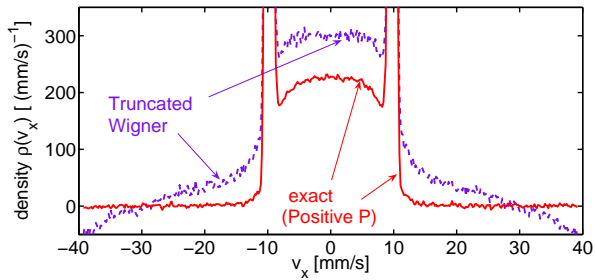}

\caption{Comparison of truncated Wigner and +P simulations, for total scattered
particle density in velocity space at a given final velocity. While
the +P results agree with experiments, the results show unphysical
negative densities and `ghost' scattering events in the approximate
truncated Wigner approach. In these simulations, space is discretized
onto a $432\times50\times50$ lattice with $k_{{\rm x,max}}=1.4\times10^{7}$/m
and $k_{{\rm y,z,max}}=6.2\times10^{6}$/m, giving $10^{6}$ modes
in total, with physical parameters given in the text. \label{fig:Comparison-of-truncated}}
\end{figure}

Collisions have been calculated for up to $10^{6}$ modes and $10^{5}$
interacting bosons\cite{prl-98-120402}. This is a clearly exponential
regime, yet the +P technique is definitely applicable. There are restrictions
on the interaction density and total time duration possible before
the sampling error is too large, but useful results are certainly
obtainable, as shown in Fig (\ref{fig:Comparison-of-truncated}).
We emphasize that sampling error, while easily estimated, is not easily
reduced at long times due to a rapid growth in the distribution variance,
which eventually makes these simulations impractical.

By comparison, with the same parameters the truncated Wigner method
clearly fails to produce physically sensible results. There is an
uncontrolled error leading to a completely unphysical depletion of
the vacuum at large relative velocity, causing apparently negative
particle densities which of course cannot occur. This depletion leads
to `ghost' particles scattering into low-velocity regions near the
original condensate velocity, which do not correspond to any real
physical events.

While the +P simulation produces physically sensible results up to
the time when sampling errors are too large to be useful, the truncated
Wigner approach is liable to generate completely unphysical behavior.
This is due to the fact that in a three-dimensional quantum field
simulation, there are a diverging number of unoccupied high-momentum
modes in the limit of high momentum cutoff. These contradict the basic
high-occupation number approximation inherent in the Wigner truncation.

In summary, the advantages and disadvantages of the +P approach are
that it treats exponentially large Hilbert spaces without mean-field
or factorization assumptions, including either unitary or non-unitary
damped evolution. No truncation of the equations of motion is required
, and there is no UV divergence at large k-value. However, for unitary
evolution, and especially for strong interactions, the sampling error
grows in time. This is intrinsic to the stochastic method, which leads
to solutions have relatively large tails. From the fundamental existence
theorem, there are known solutions that are strongly bounded with
small sampling errors, but one must use different simulation techniques
to access these solutions.

\section{Gaussian Representation}

The Gaussian phase-representation is a more general phase-space representation
than either the Wigner or positive-P method. In fact, it includes
these as special cases, and extends the phase-space idea to include
fermions \cite{Corney:2003,corney_drummond_06c}. Here we consider
a general number-conserving Gaussian operator basis, in which any
density matrix ${\rho}$ is expanded in terms of Gaussian operators,
${\Lambda}(\bm{\lambda})$, defined as exponentials of quadratic forms.
In order to define the Gaussian operators, we consider a bosonic or
fermionic quantum field with an $M$-dimensional set of mode operators
${\bm{a}}^{\dagger}\equiv\left[{a}_{1}^{\dagger},{a}_{2}^{\dagger},\ldots{a}_{M}^{\dagger}\right]$.
In the bosonic case, we can define $\delta{\bm{a}}={\bm{a}}-\bm{\alpha}$
and $\delta{\bm{a}}^{\dagger}={\bm{a}}^{\dagger}-\bm{\beta}$ as operator
displacements, where in general $\bm{\alpha}$ and $\bm{\beta}^{\dagger}$
are independent complex vectors. In the fermionic case we set these
displacements to zero. The annihilation and creation operators satisfy
(anti) commutation relations, with ($+$) for fermions and ($-$)
for bosons:

\begin{equation}
\left[{a}_{i,}{a}_{j}^{\dagger}\right]_{\pm}=\delta_{ij}.\label{eq:commutators}
\end{equation}

The expansion of the density matrix is:

\begin{equation}
{\rho}=\int P(\bm{\lambda}){\Lambda}(\bm{\lambda})d\bm{\lambda}\,\,,\label{eq:dmatrix}
\end{equation}
 where $P(\bm{\lambda})$ is the probability density over the phase-space,
$\bm{\lambda}$ is the complex vector parameter of the Gaussian representation
, $d\bm{\lambda}$ is the integration measure, and ${\Lambda}(\bm{\lambda})$
is a Gaussian operator, defined as a normally ordered exponential
of a quadratic form of annihilation and creation operators: 
\begin{equation}
{\Lambda}(\bm{\lambda})=\frac{1}{{\cal N}}{\Lambda}_{u}\left(\bm{\lambda}\right)=\frac{1}{{\cal N}}:\exp\left[-\delta{\bm{a}}^{\dagger}\underline{\bm{\mu}}\delta{\bm{a}}\right]:\label{eq:Gaussian_op}
\end{equation}
 Here, $\underline{\bm{\mu}}$ is a complex $M\times M$ matrix, so
that the representation phase space is $\bm{\lambda}=\left[\bm{\alpha},\bm{\beta},\underline{\bm{\mu}}\right]$.
${\cal N}=Tr\left[{\Lambda}_{u}(\bm{\lambda})\right]$ is a normalizing
factor, and $:\,:$ indicates normal ordering. The normalizing factor
has two forms, for bosons and fermions respectively: 
\begin{eqnarray}
{\cal N}_{b} & = & \det\left[\underline{\bm{\mu}}\right]^{-1}\nonumber \\
{\cal N}_{f} & = & \det\left[2\underline{\bm{I}}-\underline{\bm{\mu}}\right]\,.
\end{eqnarray}

The matrix $\underline{\bm{\mu}}$ is related to the stochastic Green's
function $\underline{\bm{n}}$ as:
\begin{eqnarray}
\underline{\bm{n}}_{b} & = & \underline{\bm{\mu}}^{-T}-\underline{\bm{I}}\nonumber \\
\underline{\bm{n}}_{f} & = & \left[2\underline{\bm{I}}-\underline{\bm{\mu}}\right]^{-T}\,.\label{eq:Green_sf}
\end{eqnarray}
In either case, the stochastic average of $\underline{\bm{n}}$ over
the distribution $P$ is physically a normally ordered many-body Green's
function, so that: 
\begin{equation}
\left\langle {a}_{i}^{\dagger}{a}_{j}\right\rangle =\left\langle n_{ij}+\beta_{i}\alpha_{j}\right\rangle _{P}\,.
\end{equation}

Similar methods to the positive-P approach can then be used for calculating
time-evolution. These have been used very successfully in treating,
for example, the ground state of the fermionic Hubbard model \cite{aimi_imada_07}
- a problem of much interest in condensed matter physics.

\subsection{Linear entropy}

Entropy is a measure of loss of information and entanglement in a
quantum system, and it is important to have a method of sampling a
phase-space distribution in order to estimate the entropy. Here we
show that the Gaussian phase-space method is well-suited to this type
of calculation. In particular, the linear or Renyi entropy\cite{Renyi_1960},
is defined as: 
\begin{equation}
S_{2}=-\ln Tr\left({\rho}^{2}\right)\,\,.\label{eq:quantu-Renyi}
\end{equation}
 This has similar properties to the usual logarithmic entropy, and
measures state purity, since $S_{2}=0$ for a pure state, while $S_{2}>0$
for a mixed state. The Renyi entropy in a phase-space representation
can be written using Eq. (\ref{eq:quantu-Renyi}) and the expansion
of the density matrix Eq. (\ref{eq:dmatrix}) as:

\begin{equation}
S_{2}=-\ln\iint P(\bm{\lambda})P(\bm{\lambda}')Tr\left({\Lambda}(\bm{\lambda}){\Lambda}(\bm{\lambda}')\right)\, d\bm{\lambda}d\bm{\lambda}'\,.\label{eq:Linear_S}
\end{equation}

In order to obtain an expression of the linear entropy using the Gaussian
phase-representation is necessary to evaluate the inner products of
Gaussian operators of form $Tr\left({\Lambda}(\bm{\lambda}){\Lambda}(\bm{\lambda}')\right)$\cite{ZaratePhysRevA2011}.
We emphasize here that this procedure does not give useful results
for the usual classical phase-space representations. However, we will
show that it yields remarkably simple results for both fermionic and
bosonic Gaussian representations. 

For the fermionic case, we first evaluate the trace of the inner product
of two un-normalized fermionic operators $F\left(\underline{\bm{\mu}},\underline{\bm{\nu}}\right)=Tr\left[{\Lambda}_{u}\left(\underline{\bm{\mu}}\right){\Lambda}_{u}\left(\underline{\bm{\nu}}\right)\right]$.
In this approach, the physical many-body system is treated as a distribution
over fermionic Green's functions, whose average are the observed Green's
function or correlation function. 

In order to evaluate the trace we use fermionic coherent states $\vert\bm{\alpha}\rangle$
in terms of Grassmann variables $\bm{\alpha}$\cite{Cahill1999a},
as well as the trace of an operator and the identity operator of fermionic
coherent states. After some Grassmann calculus, we obtain that the
inner product of two un-normalized fermionic Gaussian operators is:
\begin{equation}
F\left(\underline{\bm{\mu}},\underline{\bm{\nu}}\right)=\det\left[\bm{\mathrm{I}}\,+\left(\underline{\bm{I}}-\underline{\bm{\mu}}\right)\left(\underline{\bm{I}}-\underline{\bm{\nu}}\right)\right].
\end{equation}
We rewrite these expressions in terms of the normally ordered Green's
functions or correlations of the basis sets: 
\begin{equation}
n_{ij}=Tr\left[{\Lambda}(\mathbf{n}){a}_{i}^{\dagger}{a}_{j}\right].
\end{equation}
 Introducing the hole Green's functions, $\tilde{\underline{\mathbf{n}}}=\left[\underline{\bm{\mathrm{I}}}-\underline{\mathbf{n}}\right]$,
and $\tilde{\underline{\mathbf{m}}}=\left[\underline{\bm{\mathrm{I}}}-\underline{\mathbf{m}}\right]$,
we obtain the following result for the normalized inner product of
fermionic Gaussian operators: 
\begin{equation}
Tr\left[{\Lambda}(\mathbf{m}){\Lambda}(\mathbf{n})\right]=\det\left[\tilde{\underline{\mathbf{n}}}\tilde{\underline{\mathbf{m}}}+\underline{\mathbf{n}}\underline{\mathbf{m}}\right].\label{eq:Tracef}
\end{equation}

Using the result of the trace of the inner product of fermionic Gaussian
operators, Eq.$\ $(\ref{eq:Tracef}), we obtain that the expression
for the linear entropy in a Gaussian phase representation Eq.$\ $(\ref{eq:Linear_S})
is:

\begin{equation}
S_{2}=-\ln\iint P(\bm{m})P(\bm{n})\det\left[\tilde{\underline{\mathbf{n}}}\tilde{\underline{\mathbf{m}}}+\underline{\mathbf{n}}\underline{\mathbf{m}}\right]\, d\mathbf{m}d\mathbf{n}\,.\label{eq:LinearS_f}
\end{equation}

Just as in the case of fermions, we evaluate the inner product of
two un-normalized bosonic Gaussian operators, $B\left(\underline{\bm{\mu}},\underline{\bm{\nu}}\right)=Tr\left[{\Lambda}_{u}\left(\underline{\bm{\mu}}\right){\Lambda}_{u}\left(\underline{\bm{\nu}}\right)\right]$
, and after using coherent state expansions, we obtain that the inner
product of two un-normalized bosonic Gaussian operators is: 
\begin{equation}
B\left(\underline{\bm{\mu}},\underline{\bm{\nu}}\right)=\det\left[\bm{\mathrm{I}}\,-\left(\underline{\bm{\mu}}-\underline{\bm{I}}\right)\left(\underline{\bm{\nu}}-\underline{\bm{I}}\right)\right]^{-1}\label{eq:Traceb}
\end{equation}

Similar to the fermionic case, we can rewrite this expression in terms
of the stochastic Green's functions, and finally we have that:

\begin{equation}
S_{2}=\ln\iint P(\bm{m})P(\bm{n})\det\left[\underline{\bm{\mathrm{I}}}+\underline{\mathbf{n}}+\underline{\mathbf{m}}\right]\, d\mathbf{m}d\mathbf{n}\,.\label{eq:LinearS_b}
\end{equation}

In summary, using the results of the inner products of Gaussian operators,
we can obtain an expression for the linear entropy. Since this is
just an average over two independent probabilities, it is readily
calculable using sampled phase-space representations. This expression
can also be used to evaluate the entanglement of a quantum system
with a reservoir or other coupled system.

\section{Variational methods}

While the previous phase-space methods have had a long history in
physics, there is a different approach with an equally long history,
namely the use of variational techniques. This is a very simple concept,
which is that one should use an evolution equation which minimizes
the error.

The idea was first proposed by Frenkel\cite[p436]{Frenkel-Wave} and
Dirac, who developed early variational methods. In Dirac's original
approach, an effective action method was proposed for a variational
wave-function $\left|\psi\left(t\right)\right\rangle $, of the form:
\begin{equation}
\delta\Gamma=\delta\int dt\left\langle \psi\left(t\right)\right|\left[i\hbar\partial_{t}-{H}\right]\left|\psi\left(t\right)\right\rangle =0
\end{equation}

This generates a variational Schroedinger equation, which gives the
exact Schroedinger equation in the case that $\left|\psi\left(t\right)\right\rangle $
is a complete set of wave-functions. In the usual application of the
method, $\left|\psi\left(t\right)\right\rangle $ is chosen as a specific
functional form described by a small number of free parameters. The
properties of this method are that results depend on the chosen function,
and there are a small number of equations. However, one can't easily
determine errors, and of course, the method doesn't converge if $\left|\psi\left(t\right)\right\rangle $
is incomplete.

\subsection{Multiconfigurational methods}

More recent applications of the variational approach\cite{Meyer-Multidimensional}
have focused on the concept of choosing an expansion for the variational
wavefunction that is complete in some limit. These are typically sums
of individual many-body wavefunctions known as configurations, hence
the term multi-configurational approach. A common approach in the
BEC case is to construct the variational wavefunction from sums of
multiply-occupied Fock states of the form\cite{pra-77-33613}: 
\begin{equation}
\left|\psi\left(t\right)\right\rangle \ldots=\sum_{\vec{n}}C_{\vec{n}}\left(t\right){a}_{1}^{n_{1}}\ldots{a}_{M}^{n_{M}}\left|0\right\rangle \,,
\end{equation}
 where the operators ${a}_{j}^{\dagger}$ are time-dependent operators
such that: 
\begin{equation}
{a}_{j}^{\dagger}=\int d^{3}\mathbf{r}{\Psi}^{\dagger}\left(\mathbf{r}\right)\phi_{j}\left(\mathbf{r},t\right)\,.
\end{equation}

Compared to the usual variational approach, the strategy used here
is to systematically increase the Hilbert space dimension by increasing
the number of modes, with full convergence expected as $M\rightarrow\infty$
. An obvious drawback is that, since $M$ is the number of modes,
and the Hilbert space is essentially a standard Fock space, there
is clearly a potential problem with this strategy. There is an exponential
complex Hilbert space dimension as $\, M$ increases. The result of
this problem is that the technique is currently restricted to one
dimension with less than 100 particles, and relatively weak interactions.
However, given this restriction, relatively long interaction times
are possible.

\subsection{Combining phase-space and variational methods}

Given the success of phase-space approach in dealing with complexity,
an obvious question is: how can we unify variational and coherent
states phase-space methods? Such an approach would have several potential
advantages\textbf{\emph{. }}Coherent states provide description of
long-range coherence, and in principle are a complete basis. The usual
stochastic method with finite computational samples can cause a large
sampling error. However, the variational approach can be used to minimize
the error. The goal of such an approach is to combine a high degree
of complexity with relatively long time-scale evolution.

In order to describe the simplest resulting method of this type, we
recall the Wheeler-Everett multiverse concept\cite{rmp-29-454}. That
is, suppose we let $\left|\bm{\alpha}\right\rangle $ be a multi-mode
coherent state, then an $\mathcal{N}$ component superposition of
coherent states can be considered a `multiverse' - a superposition
of $\mathcal{N}$ classical worlds. This is parameterized by a coherent
`super-vector $\,\mathbf{X}\equiv\left(\bm{\alpha}^{(1)},\ldots\bm{\alpha}^{(\mathcal{N})}\right)$,
where $\bm{\alpha}=\left(\alpha_{0},\ldots\alpha_{M}\right)$ is a
multi-mode coherent amplitude, ${a}_{0}\equiv{1}$ is the unit operator,
and $\alpha_{0}$ is introduced here as a combined relative phase
and weight parameter, including the state normalization factor. The
corresponding quantum state defined as: 
\begin{align}
\left|\psi\left(t\right)\right\rangle  & =\,\sum_{n=1}^{\mathcal{N}}e^{\bm{\alpha}^{(n)}\left(t\right)\cdot{\mathbf{a}}^{\dagger}}\left|0\right\rangle \label{eq:multiverse}
\end{align}
 This is a constrained multiverse, with a fixed number of copies,
shown schematically in Fig (\ref{fig:Multiverse}). One cannot keep
track of \emph{all} quantum universes! It is also, in quantum mechanical
terms, a superposition of non-orthogonal coherent states. We find
that the use of a variational principle introduces interactions between
coherent amplitudes. The advantage is a much lower sampling error
compared to independent stochastic evolution.

{ 
\begin{figure}
\includegraphics[width=0.9\columnwidth]{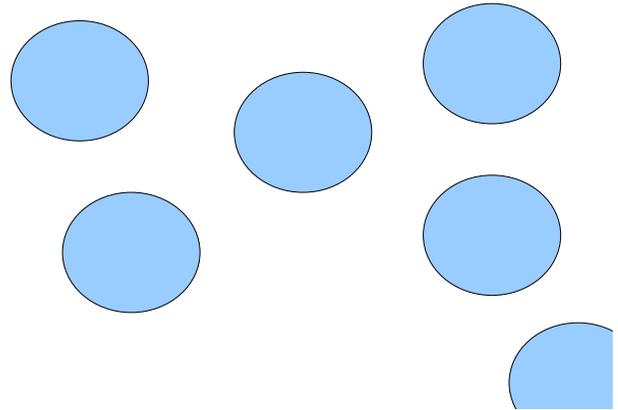} \caption{What does the multiverse look like? Here we give an illustration of
a quantum `universe' as a superposition of coherent states, illustrated
by the blue circles. Each blue circle represents a different multi-mode
coherent state $\bm{\alpha}^{(n)}\left(t\right)$, which has an intrinsic
uncertainty. The whole quantum state or `universe' is a superposition
of $\mathcal{N}$ coherent states with different phases and amplitudes.\label{fig:Multiverse}}
\end{figure}

\subsection{Phase-space variational equations}

Having introduced the concept, we now wish to derive the resulting
equations\cite{jcp-128-044116}. Suppose the Hamiltonian is $\, H\left({\mathbf{a}}^{\dagger},{\mathbf{a}}\right)$,
with an exact wavefunction $\left|\tilde{\psi}\right\rangle $, and
a time-interval $\Delta t=t-t_{0}$, then clearly: 
\begin{equation}
\,\left|\tilde{\psi}\left(t\right)\right\rangle =e^{-i{H}\Delta t}\left|\psi\left(t_{0}\right)\right\rangle \,\,.
\end{equation}
 We wish to minimize local propagation error in calculating the computational
result$\left|\psi\left(t\right)\right\rangle $, so that:

\begin{equation}
\delta\mathcal{E}=\,\delta\left\Vert \left|\psi\left(t\right)\right\rangle -e^{-i{H}\Delta t}\left|\psi\left(t_{0}\right)\right\rangle \right\Vert ^{2}=0.
\end{equation}
 Expanding to first order and taking the limit of $\Delta t=t-t_{0}\rightarrow0$,
gives:

\begin{equation}
\Re\left\langle \delta\psi\right|\left[\partial_{t}+i{H}\right]\left|\psi\right\rangle =0\,.
\end{equation}
 We introduce $\left|\Delta\psi\right\rangle =\left|\psi\left(t\right)\right\rangle -\left|\psi\left(t_{0}\right)\right\rangle $
and linearize in the coherent state parameters, so that: 
\begin{equation}
\left|\Delta\psi\right\rangle =\Delta X_{\mu}\frac{\partial\left|\psi\right\rangle }{\partial X_{\mu}}
\end{equation}
 The variational principle then leads to a differential equation for
the coherent parameters $\mathbf{X}$, of form: 
\begin{equation}
\,\partial_{t}\mathbf{X}=-i\mathcal{\boldsymbol{V}}^{-1}\mathcal{\mathcal{\mathbf{H}}}\,.
\end{equation}

This means that there are now \textbf{\emph{$P\equiv\mathcal{N}(M+1)$}}
differential equations to solve, with a combined index $\mu,$ where
the main terms are 
\begin{itemize}
\item The variational matrix: $\,\mathcal{V}_{\mu\nu}=\left\{ \frac{\partial}{\partial X_{\mu}}\left\langle \psi\right|\right\} \,\frac{\partial}{\partial X_{\nu}}\left|\psi\right\rangle $ 
\item The H-vector: $\,\mathcal{H}_{\mu}=\left\{ \frac{\partial}{\partial X_{\mu}}\left\langle \psi\right|\right\} \,{H}\left|\psi\left(t\right)\right\rangle .$ 
\end{itemize}
However, a numerical problem must be treated, which is that $\mathcal{V}_{\mu\nu}$
is generally not invertible, owing to the existence of multiple minima.
We can solve this iteratively, in terms of the parameter change $\Delta\mathbf{X}^{[p]}$,
where we set $\Delta\mathbf{X}^{[0]}=0$ initially and iterate until
a stable solution is reached. This is similar to the Tikhonov variational
method\cite{Tikhonov-Nonlinear}. In detail, suppose we have a set
of variational parameters $\mathbf{X}\left(t_{0}\right)$, at time
$t=t_{0}$. We evaluate the variational matrix and Hamiltonian vector
a midpoint $t_{0}+\Delta t/2$, by iterating so that $\mathbf{X}^{[p-1]}=\mathbf{X}\left(t_{0}\right)+\Delta\mathbf{X}^{[p-1]}$,
and setting the change in $\mathbf{X}$ to $\Delta\mathbf{X}^{[p]}$,
where: 
\begin{align}
\Delta\mathbf{X}^{[p]} & =\Delta\mathbf{X}^{[p-1]}+\left[\mathcal{\mathbf{\mathcal{V}}}^{[p-1]}+i\lambda\mathbf{I}\right]^{-1}\times\nonumber \\
 & \times\left[-i\Delta t\mathcal{H}^{[p-1]}/2-\mathcal{\mathbf{\mathcal{V}}}^{[p-1]}\Delta\mathbf{X}^{[p-1]}\right]
\end{align}

The final step is to propagate to $t_{0}+\Delta t$ by setting: 
\begin{equation}
\mathbf{X}\left(t_{0}+\Delta t\right)=\mathbf{X}\left(t_{0}\right)+2\Delta\mathbf{X}^{[p]}\,,
\end{equation}
in order to move to the next step in time. This method leads to stable
equations, with no inversion problems. 

In the case of a coherent state expansion, we define an energy matrix:
\begin{equation}
H^{(m,n)}=H\left(\bm{\alpha}^{(m)*},\bm{\alpha}^{(n)}\right)\,,
\end{equation}
 a reduced amplitude: 
\begin{equation}
\,\tilde{\alpha}_{k}^{(m)}=\delta_{k0}+\left[1-\delta_{k0}\right]\alpha_{k}^{(m)},\,
\end{equation}
 and an inner product: 
\begin{equation}
\,\rho^{(mn)}=\exp\left[\alpha_{0}^{(m)*}+\alpha_{0}^{(m)}+\sum_{k>0}\alpha_{k}^{(m)*}\alpha_{k}^{(n)}\right]\,.
\end{equation}

The variational matrix definitions are then: 
\begin{align}
V_{kl}^{(mn)} & \,=\,\left[\left[1-\delta_{k0}\right]\delta_{kl}+\tilde{\alpha}_{l}^{(m)*}\tilde{\alpha}_{k}^{(n)}\right]\rho^{(mn)}\nonumber \\
H_{k}^{(m)} & \,=\,\sum_{n}\left[\frac{\partial H^{(m,n)}}{\partial\alpha_{k}^{(m)*}}+H^{(m,n)}\tilde{\alpha}_{k}^{(n)}\right]\rho^{(mn)}\,\,.
\end{align}

~

For example, in the case of a single variational term, one finds equations
equivalent to the Gross-Pitaevskii mean-field equation, with an additional
phase evolution for $\alpha_{0}$:

\begin{eqnarray}
\,\partial_{t}\alpha_{0} & \,= & \,-i\left[H-\bm{\alpha}^{\dagger}\cdot\partial H/\partial\bm{\alpha}^{\dagger}\right]\nonumber \\
\,\partial_{t}\bm{\alpha} & \,= & \,-i\partial H/\partial\bm{\alpha}^{\dagger}\,\,.
\end{eqnarray}
For a linear Hamiltonian: 
\begin{equation}
\,{H}={\mathbf{a}}^{\dagger}\bm{\omega}{\mathbf{a}}\,
\end{equation}
 one finds that each coherent term evolves independently of each other
term, giving: 
\begin{eqnarray}
\,\partial_{t}\alpha_{0}^{(n)} & \,= & \,0\nonumber \\
\,\partial_{t}\bm{\alpha}^{(n)} & \,= & \,-i\bm{\omega}\bm{\alpha}^{(n)}\,\,.
\end{eqnarray}
 In this case each linear `universe' is decoupled from the others;
this is an exact result, and no approximations are required.

\subsection{Recurrences in the anharmonic oscillator}

Finally, we consider the case of a quantum anharmonic oscillator,
which describes local S-wave scattering interactions at a single lattice
site, so that: 
\begin{equation}
\,{H}={a}^{\dagger2}{a}^{2}
\end{equation}
 This is an extremely strong test of coherent state expansion\textbf{
}methods. From an initial coherent state the quantum evolution generates
Schroedinger cat superpositions, with complete recurrences known to
occur analytically\cite{pra-33-674}, as well as experimentally \cite{Greiner_02}.
Variational results showing almost error-free dynamics throughout
the Schroedinger-cat regime of $t=\pi$, and a complete recurrence
at $t=2\pi$ are shown in the figures. For these numerical results,
we used $\lambda=10^{-4}$ to control the matrix inversion, with four
iterations of the variational equations at each time point.

\begin{figure}
\includegraphics[width=0.9\columnwidth]{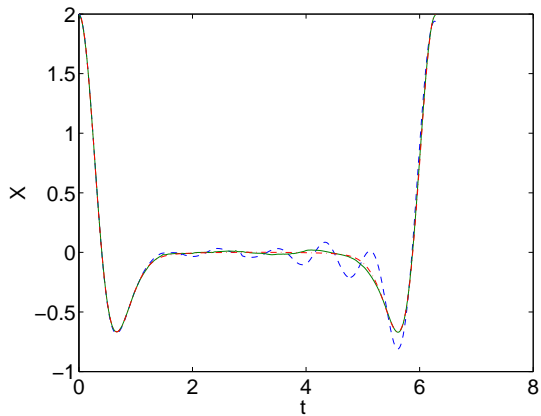}\caption{Anharmonic recurrence, x quadrature. Blue dashed line has $8$ components,
green solid line has $16$ components. This is almost indistinguishable
from the the exact solution, which is a red dashed line. \label{fig:Anharmonic-recurrence-x}}
\end{figure}

Here we define quadrature variables, 
\begin{equation}
X=\left\langle {a}+{a}^{\dagger}\right\rangle /2
\end{equation}

and 
\begin{equation}
Y=\left\langle {a}-{a}^{\dagger}\right\rangle /(2i\mbox{})
\end{equation}

The exact result, given an initial coherent state, is known to have
recurrences in both quadratures, as shown in the graphs.

\begin{figure}
\includegraphics[width=0.9\columnwidth]{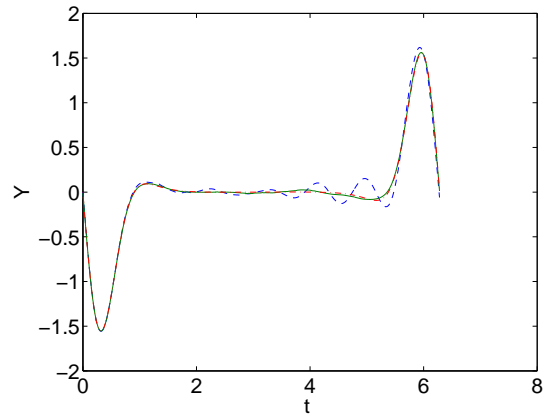}\caption{Anharmonic recurrence, y quadrature. Parameters as in the previous
figure. The blue dashed line with small oscillations is not quite
converged, while the solid green line shows excellent convergence
to the exact result. \label{fig:Anharmonic-recurrence-y}}
\end{figure}

While it is simple enough to be analytically soluble, this Hamiltonian
would lead to large errors with stochastic phase-space techniques
over time-scales comparable to the recurrence time. As we show in
Figs (\ref{fig:Anharmonic-recurrence-x}) and (\ref{fig:Anharmonic-recurrence-y}),
the use of a variational method allows us to track the full recurrence,
on timescales where any previous stochastic phase-space method would
give very large errors \cite{Deuar:2002,deuar_drummond_06,dowling_drummond_05}.
Variational convergence is rapid, with only small errors visible using
$\mathcal{N}=8$ components. These are almost completely eliminated
by using $\mathcal{N}=16$ coherent components. This illustrative
example is very simple, and indeed the case treated here can be solved
exactly using analytic techniques. However, it does illustrate the
utility of variational methods in reducing sampling error in phase-space
simulations.

\section{Outlook and summary}

In summary, there are a growing number of experiments in ultra-cold
atomic physics that probe the world of quantum dynamics. Simple, exact
results are only possible with a small number of interacting modes.
Larger, complex quantum systems require new techniques to handle the
issue of exponential complexity. While phase-space representations
using non-orthogonal basis sets can can treat highly complex problems,
there is often a problem with truncations (in the Wigner case) or
sampling errors that grow in time (in the positive-P case). 

We have shown that there are newer techniques that have much promise.
Gaussian phase-space methods are much more general, and can treat
new issues like fermions and entropy calculations. Finally, we have
derived a Tikhonov-based variational approach which is shown to dramatically
reduce sampling errors in a case which is known to be challenging
to handle with phase-space methods. This approach gives exact results
for linear couplings. It has lower sampling errors than the stochastic
+P method, as well as reduced variational complexity issues compared
to multi-configurational approaches in a Fock space basis. While the
techniques given here are only preliminary, this hybrid variational
and phase-space approach appears promising for future developments.

\bibliographystyle{apsrev4-1}
\bibliography{frontiers,phase-space}

%Merlin.mbs v4.21 2009-07-09.
\begin{thebibliography}{10}%
\makeatletter
\providecommand \@ifxundefined [1]{%
 \ifx #1\undefined \expandafter \@firstoftwo
 \else \expandafter \@secondoftwo
\fi
}%
\providecommand \@ifnum [1]{%
 \ifnum #1\expandafter \@firstoftwo
 \else \expandafter \@secondoftwo
\fi
}%
\providecommand \enquote [1]{``#1''}%
\providecommand \bibnamefont  [1]{#1}%
\providecommand \bibfnamefont [1]{#1}%
\providecommand \citenamefont [1]{#1}%
\providecommand\href[0]{\@sanitize\@href}%
\providecommand\@href[1]{\endgroup\@@startlink{#1}\endgroup\@@href}%
\providecommand\@@href[1]{#1\@@endlink}%
\providecommand \@sanitize [0]{\begingroup\catcode`\&12\catcode`\#12\relax}%
\@ifxundefined \pdfoutput {\@firstoftwo}{%
 \@ifnum{\z@=\pdfoutput}{\@firstoftwo}{\@secondoftwo}%
}{%
 \providecommand\@@startlink[1]{\leavevmode\special{html:<a href="#1">}}%
 \providecommand\@@endlink[0]{\special{html:</a>}}%
}{%
 \providecommand\@@startlink[1]{%
  \leavevmode
  \pdfstartlink
   attr{/Border[0 0 1 ]/H/I/C[0 1 1]}%
   user{/Subtype/Link/A<</Type/Action/S/URI/URI(#1)>>}%
  \relax
 }%
 \providecommand\@@endlink[0]{\pdfendlink}%
}%
\providecommand \url  [0]{\begingroup\@sanitize \@url }%
\providecommand \@url [1]{\endgroup\@href {#1}{\urlprefix}}%
\providecommand \urlprefix [0]{URL }%
\providecommand \Eprint[0]{\href }%
\@ifxundefined \urlstyle {%
  \providecommand \doi [1]{doi:\discretionary{}{}{}#1}%
}{%
  \providecommand \doi [0]{doi:\discretionary{}{}{}\begingroup
  \urlstyle{rm}\Url }%
}%
\providecommand \doibase [0]{http://dx.doi.org/}%
\providecommand \Doi[1]{\href{\doibase#1}}%
\providecommand \bibAnnote [3]{%
  \BibitemShut{#1}%
  \begin{quotation}\noindent
    \textsc{Key:}\ #2\\\textsc{Annotation:}\ #3%
  \end{quotation}%
}%
\providecommand \bibAnnoteFile [2]{%
  \IfFileExists{#2}{\bibAnnote {#1} {#2} {\input{#2}}}{}%
}%
\providecommand \typeout [0]{\immediate \write \m@ne }%
\providecommand \selectlanguage [0]{\@gobble}%
\providecommand \bibinfo [0]{\@secondoftwo}%
\providecommand \bibfield [0]{\@secondoftwo}%
\providecommand \translation [1]{[#1]}%
\providecommand \BibitemOpen[0]{}%
\providecommand \bibitemStop [0]{}%
\providecommand \bibitemNoStop [0]{.\EOS\space}%
\providecommand \EOS [0]{\spacefactor3000\relax}%
\providecommand \BibitemShut [1]{\csname bibitem#1\endcsname}%
%</preamble>
\bibitem{prl-84-5029}%
  \BibitemOpen
  \bibfield{author}{%
  \bibinfo {author} {\bibfnamefont{D.}~\bibnamefont{Heinzen}}, \bibinfo
  {author} {\bibfnamefont{R.}~\bibnamefont{Wynar}}, \bibinfo {author}
  {\bibfnamefont{P.~D.}\ \bibnamefont{Drummond}},\ and\ \bibinfo {author}
  {\bibfnamefont{K.~V.}\ \bibnamefont{Kheruntsyan}},\ }%
  \bibfield{journal}{%
  \Doi{10.1103/PhysRevLett.84.5029}{\bibinfo {journal} {Physical Review
  Letters}}\ }%
  \textbf{\bibinfo {volume} {84}},\ \bibinfo {pages} {5029} (\bibinfo {month}
  {May}\ \bibinfo {year} {2000}),\ ISSN \bibinfo {issn} {0031-9007},\
  \url{http://link.aps.org/doi/10.1103/PhysRevLett.84.5029}%
  \bibAnnoteFile{NoStop}{prl-84-5029}%
\bibitem{nat-417-529}%
  \BibitemOpen
  \bibfield{author}{%
  \bibinfo {author} {\bibfnamefont{E.~A.}\ \bibnamefont{Donley}}, \bibinfo
  {author} {\bibfnamefont{N.~R.}\ \bibnamefont{Claussen}}, \bibinfo {author}
  {\bibfnamefont{S.~T.}\ \bibnamefont{Thompson}},\ and\ \bibinfo {author}
  {\bibfnamefont{C.~E.}\ \bibnamefont{Wieman}},\ }%
  \bibfield{journal}{%
  \Doi{10.1038/417529a}{\bibinfo {journal} {Nature}}\ }%
  \textbf{\bibinfo {volume} {417}},\ \bibinfo {pages} {529} (\bibinfo {month}
  {May}\ \bibinfo {year} {2002}),\ ISSN \bibinfo {issn} {0028-0836},\
  \url{http://www.ncbi.nlm.nih.gov/pubmed/12037562}%
  \bibAnnoteFile{NoStop}{nat-417-529}%
\bibitem{sci-310-1513}%
  \BibitemOpen
  \bibfield{author}{%
  \bibinfo {author} {\bibfnamefont{A.~E.}\ \bibnamefont{Leanhardt}}, \bibinfo
  {author} {\bibfnamefont{T.~A.}\ \bibnamefont{Pasquini}}, \bibinfo {author}
  {\bibfnamefont{M.}~\bibnamefont{Saba}}, \bibinfo {author}
  {\bibfnamefont{A.}~\bibnamefont{Schirotzek}}, \bibinfo {author}
  {\bibfnamefont{Y.}~\bibnamefont{Shin}}, \bibinfo {author}
  {\bibfnamefont{D.}~\bibnamefont{Kielpinski}}, \bibinfo {author}
  {\bibfnamefont{D.~E.}\ \bibnamefont{Pritchard}},\ and\ \bibinfo {author}
  {\bibfnamefont{W.}~\bibnamefont{Ketterle}},\ }%
  \bibfield{journal}{%
  \Doi{10.1126/science.1088827}{\bibinfo {journal} {Science (New York, N.Y.)}}\
  }%
  \textbf{\bibinfo {volume} {301}},\ \bibinfo {pages} {1513} (\bibinfo {month}
  {Sep.}\ \bibinfo {year} {2003}),\ ISSN \bibinfo {issn} {1095-9203},\
  \url{http://www.sciencemag.org/content/301/5639/1513.abstract}%
  \bibAnnoteFile{NoStop}{sci-310-1513}%
\bibitem{Datta-Electronic}%
  \BibitemOpen
  \bibfield{author}{%
  \bibinfo {author} {\bibfnamefont{S.}~\bibnamefont{Datta}},\ }%
  \emph{\bibinfo {title} {{Electronic Transport in Mesoscopic Systems}}}\
  (\bibinfo {publisher} {Cambridge University Press},\ \bibinfo {address}
  {Cambridge},\ \bibinfo {year} {1995})\ ISBN \bibinfo {isbn} {0521599431},\
  \url{http://www.amazon.com/Electronic-Mesoscopic-Semiconductor-Microelectron%
ic-Engineering/dp/0521599431/}%
  \bibAnnoteFile{NoStop}{Datta-Electronic}%
\bibitem{Metcalf-Laser}%
  \BibitemOpen
  \bibfield{author}{%
  \bibinfo {author} {\bibfnamefont{H.~J.}\ \bibnamefont{Metcalf}}\ and\
  \bibinfo {author} {\bibfnamefont{P.}~\bibnamefont{van~der Straten}},\ }%
  \emph{\bibinfo {title} {{Laser Cooling and Trapping}}}\ (\bibinfo {publisher}
  {Springer},\ \bibinfo {year} {1999})\ ISBN \bibinfo {isbn} {0387987282},\ p.\
  \bibinfo {pages} {339},\
  \url{http://www.amazon.com/Cooling-Trapping-Graduate-Contemporary-Physics/dp%
/0387987282}%
  \bibAnnote{NoStop}{Metcalf-Laser}{Bose-Einstein condensate textbook.}%
\bibitem{sci-310-648}%
  \BibitemOpen
  \bibfield{author}{%
  \bibinfo {author} {\bibfnamefont{M.}~\bibnamefont{Schellekens}}, \bibinfo
  {author} {\bibfnamefont{R.}~\bibnamefont{Hoppeler}}, \bibinfo {author}
  {\bibfnamefont{A.}~\bibnamefont{Perrin}}, \bibinfo {author}
  {\bibfnamefont{J.~V.}\ \bibnamefont{Gomes}}, \bibinfo {author}
  {\bibfnamefont{D.}~\bibnamefont{Boiron}}, \bibinfo {author}
  {\bibfnamefont{A.}~\bibnamefont{Aspect}},\ and\ \bibinfo {author}
  {\bibfnamefont{C.~I.}\ \bibnamefont{Westbrook}},\ }%
  \bibfield{journal}{%
  \Doi{10.1126/science.1118024}{\bibinfo {journal} {Science (New York, N.Y.)}}\
  }%
  \textbf{\bibinfo {volume} {310}},\ \bibinfo {pages} {648} (\bibinfo {month}
  {Oct.}\ \bibinfo {year} {2005}),\ ISSN \bibinfo {issn} {1095-9203},\
  \url{http://www.sciencemag.org/content/310/5748/648.abstract}%
  \bibAnnoteFile{NoStop}{sci-310-648}%
\bibitem{sci-292-461}%
  \BibitemOpen
  \bibfield{author}{%
  \bibinfo {author} {\bibfnamefont{A.}~\bibnamefont{Robert}}, \bibinfo {author}
  {\bibfnamefont{O.}~\bibnamefont{Sirjean}}, \bibinfo {author}
  {\bibfnamefont{A.}~\bibnamefont{Browaeys}}, \bibinfo {author}
  {\bibfnamefont{J.}~\bibnamefont{Poupard}}, \bibinfo {author}
  {\bibfnamefont{S.}~\bibnamefont{Nowak}}, \bibinfo {author}
  {\bibfnamefont{D.}~\bibnamefont{Boiron}}, \bibinfo {author}
  {\bibfnamefont{C.~I.}\ \bibnamefont{Westbrook}},\ and\ \bibinfo {author}
  {\bibfnamefont{A.}~\bibnamefont{Aspect}},\ }%
  \bibfield{journal}{%
  \Doi{10.1126/science.1060622}{\bibinfo {journal} {Science (New York, N.Y.)}}\
  }%
  \textbf{\bibinfo {volume} {292}},\ \bibinfo {pages} {461} (\bibinfo {month}
  {Apr.}\ \bibinfo {year} {2001}),\ ISSN \bibinfo {issn} {0036-8075},\
  \url{http://www.sciencemag.org/content/292/5516/461.abstract}%
  \bibAnnoteFile{NoStop}{sci-292-461}%
\bibitem{prl-89-020401}%
  \BibitemOpen
  \bibfield{author}{%
  \bibinfo {author} {\bibfnamefont{J.~M.}\ \bibnamefont{Vogels}}, \bibinfo
  {author} {\bibfnamefont{K.}~\bibnamefont{Xu}},\ and\ \bibinfo {author}
  {\bibfnamefont{W.}~\bibnamefont{Ketterle}},\ }%
  \bibfield{journal}{%
  \Doi{10.1103/PhysRevLett.89.020401}{\bibinfo {journal} {Physical Review
  Letters}}\ }%
  \textbf{\bibinfo {volume} {89}},\ \bibinfo {pages} {020401} (\bibinfo {month}
  {Mar.}\ \bibinfo {year} {2002}),\
  \Eprint{http://arxiv.org/abs/0203286}{arXiv:0203286 [cond-mat]},\
  \url{http://arxiv.org/abs/cond-mat/0203286}%
  \bibAnnoteFile{NoStop}{prl-89-020401}%
\bibitem{Berezin-Method}%
  \BibitemOpen
  \bibfield{author}{%
  \bibinfo {author} {\bibfnamefont{F.~A.}\ \bibnamefont{Berezin}},\ }%
  \emph{\bibinfo {title} {{The Method of Second Quantization}}}\ (\bibinfo
  {publisher} {Academic Press},\ \bibinfo {address} {New York},\ \bibinfo
  {year} {1966})\ ISBN \bibinfo {isbn} {0120894505},\
  \url{http://www.amazon.com/Method-Second-Quantization-F-Berezin/dp/012089450%
5}%
  \bibAnnote{NoStop}{Berezin-Method}{Available in U Melbourne physics library,
  530.12 B492 Translated from the Russian. How does one cite it?}%
\bibitem{drummond_carter_87}%
  \BibitemOpen
  \bibfield{author}{%
  \bibinfo {author} {\bibfnamefont{P.~D.}\ \bibnamefont{Drummond}}\ and\
  \bibinfo {author} {\bibfnamefont{S.~J.}\ \bibnamefont{Carter}},\ }%
  \bibfield{journal}{%
  \bibinfo {journal} {J. Opt. Soc. Am. B}\ }%
  \textbf{\bibinfo {volume} {4}},\ \bibinfo {pages} {1565} (\bibinfo {year}
  {1987})%
  \bibAnnoteFile{NoStop}{drummond_carter_87}%
\bibitem{prl-82-2022}%
  \BibitemOpen
  \bibfield{author}{%
  \bibinfo {author} {\bibfnamefont{D.-I.}\ \bibnamefont{Choi}}\ and\ \bibinfo
  {author} {\bibfnamefont{Q.}~\bibnamefont{Niu}},\ }%
  \bibfield{journal}{%
  \Doi{10.1103/PhysRevLett.82.2022}{\bibinfo {journal} {Physical Review
  Letters}}\ }%
  \textbf{\bibinfo {volume} {82}},\ \bibinfo {pages} {2022} (\bibinfo {month}
  {Mar.}\ \bibinfo {year} {1999}),\ ISSN \bibinfo {issn} {0031-9007},\
  \url{http://link.aps.org/doi/10.1103/PhysRevLett.82.2022}%
  \bibAnnoteFile{NoStop}{prl-82-2022}%
\bibitem{jpb-35-3095}%
  \BibitemOpen
  \bibfield{author}{%
  \bibinfo {author} {\bibfnamefont{J.~H.}\ \bibnamefont{Denschlag}}, \bibinfo
  {author} {\bibfnamefont{J.~E.}\ \bibnamefont{Simsarian}}, \bibinfo {author}
  {\bibfnamefont{H.}~\bibnamefont{Haeffner}}, \bibinfo {author}
  {\bibfnamefont{C.}~\bibnamefont{McKenzie}}, \bibinfo {author}
  {\bibfnamefont{A.}~\bibnamefont{Browaeys}}, \bibinfo {author}
  {\bibfnamefont{D.}~\bibnamefont{Cho}}, \bibinfo {author}
  {\bibfnamefont{K.}~\bibnamefont{Helmerson}}, \bibinfo {author}
  {\bibfnamefont{S.~L.}\ \bibnamefont{Rolston}},\ and\ \bibinfo {author}
  {\bibfnamefont{W.~D.}\ \bibnamefont{Phillips}},\ }%
  \bibfield{journal}{%
  \Doi{10.1088/0953-4075/35/14/307}{\bibinfo {journal} {Journal of Physics B:
  Atomic, Molecular and Optical Physics}}\ }%
  \textbf{\bibinfo {volume} {35}},\ \bibinfo {pages} {3095} (\bibinfo {month}
  {Jun.}\ \bibinfo {year} {2002}),\ ISSN \bibinfo {issn} {09534075},\
  \Eprint{http://arxiv.org/abs/0206063}{arXiv:0206063 [cond-mat]},\
  \url{http://stacks.iop.org/0953-4075/35/i=14/a=307?key=crossref.143851fbd4f7%
95110e47d745ef41278c http://arxiv.org/abs/cond-mat/0206063}%
  \bibAnnoteFile{NoStop}{jpb-35-3095}%
\bibitem{rsa-276-238}%
  \BibitemOpen
  \bibfield{author}{%
  \bibinfo {author} {\bibfnamefont{J.}~\bibnamefont{Hubbard}},\ }%
  \bibfield{journal}{%
  \Doi{10.1098/rspa.1963.0204}{\bibinfo {journal} {Proceedings of the Royal
  Society A: Mathematical, Physical and Engineering Sciences}}\ }%
  \textbf{\bibinfo {volume} {276}},\ \bibinfo {pages} {238} (\bibinfo {month}
  {Nov.}\ \bibinfo {year} {1963}),\ ISSN \bibinfo {issn} {1364-5021},\
  \url{http://rspa.royalsocietypublishing.org/cgi/doi/10.1098/rspa.1963.0204}%
  \bibAnnote{NoStop}{rsa-276-238}{Swinburne doesn't have it!}%
\bibitem{itp-21-467}%
  \BibitemOpen
  \bibfield{author}{%
  \bibinfo {author} {\bibfnamefont{R.~P.}\ \bibnamefont{Feynman}},\ }%
  \bibfield{journal}{%
  \Doi{10.1007/BF02650179}{\bibinfo {journal} {International Journal of
  Theoretical Physics}}\ }%
  \textbf{\bibinfo {volume} {21}},\ \bibinfo {pages} {467} (\bibinfo {month}
  {Jun.}\ \bibinfo {year} {1982}),\ ISSN \bibinfo {issn} {0020-7748},\
  \url{http://www.springerlink.com/index/10.1007/BF02650179
  http://www.phy.mtu.edu/~sgowtham/PH4390/Week\_02/IJTP\_v21\_p467\_y1982.pdf}%
  \bibAnnoteFile{NoStop}{itp-21-467}%
\bibitem{EPR}%
  \BibitemOpen
  \bibfield{author}{%
  \bibinfo {author} {\bibfnamefont{Q.~Y.}\ \bibnamefont{He}}, \bibinfo {author}
  {\bibfnamefont{M.~D.}\ \bibnamefont{Reid}}, \bibinfo {author}
  {\bibfnamefont{T.~G.}\ \bibnamefont{Vaughan}}, \bibinfo {author}
  {\bibfnamefont{C.}~\bibnamefont{Gross}}, \bibinfo {author}
  {\bibfnamefont{M.}~\bibnamefont{Oberthaler}},\ and\ \bibinfo {author}
  {\bibfnamefont{P.~D.}\ \bibnamefont{Drummond}},\ }%
  \bibfield{journal}{%
  \bibinfo {journal} {Phys. Rev. Lett.}\ }%
  \textbf{\bibinfo {volume} {106}} (\bibinfo {year} {2011})%
  \bibAnnoteFile{NoStop}{EPR}%
\bibitem{olsen}%
  \BibitemOpen
  \bibfield{author}{%
  \bibinfo {author} {\bibfnamefont{T.~J.}\ \bibnamefont{Haigh}}, \bibinfo
  {author} {\bibfnamefont{A.~J.}\ \bibnamefont{Ferris}},\ and\ \bibinfo
  {author} {\bibfnamefont{M.~K.}\ \bibnamefont{Olsen}},\ }%
  \bibfield{journal}{%
  \bibinfo {journal} {Optics Communications}\ }%
  \textbf{\bibinfo {volume} {283}},\ \bibinfo {pages} {3540} (\bibinfo {year}
  {2010})%
  \bibAnnoteFile{NoStop}{olsen}%
\bibitem{fiberexperimentNature}%
  \BibitemOpen
  \bibfield{author}{%
  \bibinfo {author} {\bibfnamefont{P.~D.}\ \bibnamefont{Drummond}}, \bibinfo
  {author} {\bibfnamefont{R.~M.}\ \bibnamefont{Shelby}}, \bibinfo {author}
  {\bibfnamefont{S.~R.}\ \bibnamefont{Friberg}},\ and\ \bibinfo {author}
  {\bibfnamefont{Y.}~\bibnamefont{Yamamoto}},\ }%
  \bibfield{journal}{%
  \bibinfo {journal} {Nature}\ }%
  \textbf{\bibinfo {volume} {365}},\ \bibinfo {pages} {307} (\bibinfo {year}
  {1993})%
  \bibAnnoteFile{NoStop}{fiberexperimentNature}%
\bibitem{FiberEntangle}%
  \BibitemOpen
  \bibfield{author}{%
  \bibinfo {author} {\bibfnamefont{V.}~\bibnamefont{Giovannetti}}, \bibinfo
  {author} {\bibfnamefont{S.}~\bibnamefont{Mancini}}, \bibinfo {author}
  {\bibfnamefont{D.}~\bibnamefont{Vitali}},\ and\ \bibinfo {author}
  {\bibfnamefont{P.}~\bibnamefont{Tombesi}},\ }%
  \bibfield{journal}{%
  \bibinfo {journal} {Phys. Rev. A}\ }%
  \textbf{\bibinfo {volume} {67}},\ \bibinfo {pages} {022320} (\bibinfo {year}
  {2003})%
  \bibAnnoteFile{NoStop}{FiberEntangle}%
\bibitem{FiberEntangleLeuchs}%
  \BibitemOpen
  \bibfield{author}{%
  \bibinfo {author} {\bibfnamefont{R.}~\bibnamefont{Dong}}, \bibinfo {author}
  {\bibfnamefont{J.}~\bibnamefont{Heersink}}, \bibinfo {author}
  {\bibfnamefont{J.-I.}\ \bibnamefont{Yoshikawa}}, \bibinfo {author}
  {\bibfnamefont{O.}~\bibnamefont{Glockl}}, \bibinfo {author}
  {\bibfnamefont{U.~L.}\ \bibnamefont{Andersen}},\ and\ \bibinfo {author}
  {\bibfnamefont{G.}~\bibnamefont{Leuchs}},\ }%
  \bibfield{journal}{%
  \bibinfo {journal} {New J. Phys.}\ }%
  \textbf{\bibinfo {volume} {9}},\ \bibinfo {pages} {410} (\bibinfo {year}
  {2007})%
  \bibAnnoteFile{NoStop}{FiberEntangleLeuchs}%
\bibitem{BowenPolarizationEnt}%
  \BibitemOpen
  \bibfield{author}{%
  \bibinfo {author} {\bibfnamefont{W.~P.}\ \bibnamefont{Bowen}}, \bibinfo
  {author} {\bibfnamefont{N.}~\bibnamefont{Treps}}, \bibinfo {author}
  {\bibfnamefont{R.}~\bibnamefont{Schnabel}},\ and\ \bibinfo {author}
  {\bibfnamefont{P.~K.}\ \bibnamefont{Lam}},\ }%
  \bibfield{journal}{%
  \bibinfo {journal} {Phys. Rev. Lett.}\ }%
  \textbf{\bibinfo {volume} {89}},\ \bibinfo {pages} {253601} (\bibinfo {year}
  {2002})%
  \bibAnnoteFile{NoStop}{BowenPolarizationEnt}%
\bibitem{Gardiner-Quantum}%
  \BibitemOpen
  \bibfield{author}{%
  \bibinfo {author} {\bibfnamefont{C.~W.}\ \bibnamefont{Gardiner}}\ and\
  \bibinfo {author} {\bibfnamefont{P.}~\bibnamefont{Zoller}},\ }%
  \emph{\bibinfo {title} {{Quantum Noise}}},\ \bibinfo {edition} {2nd}\ ed.\
  (\bibinfo {publisher} {Springer-Verlag},\ \bibinfo {address} {Berlin},\
  \bibinfo {year} {2000})\ ISBN \bibinfo {isbn} {3-540-66571-4}%
  \bibAnnoteFile{NoStop}{Gardiner-Quantum}%
\bibitem{wigner_32}%
  \BibitemOpen
  \bibfield{author}{%
  \bibinfo {author} {\bibfnamefont{E.~P.}\ \bibnamefont{Wigner}},\ }%
  \bibfield{journal}{%
  \bibinfo {journal} {Phys. Rev.}\ }%
  \textbf{\bibinfo {volume} {40}},\ \bibinfo {pages} {749} (\bibinfo {year}
  {1932})%
  \bibAnnoteFile{NoStop}{wigner_32}%
\bibitem{glauber_63}%
  \BibitemOpen
  \bibfield{author}{%
  \bibinfo {author} {\bibfnamefont{R.~J.}\ \bibnamefont{Glauber}},\ }%
  \bibfield{journal}{%
  \bibinfo {journal} {Phys. Rev.}\ }%
  \textbf{\bibinfo {volume} {131}},\ \bibinfo {pages} {2766} (\bibinfo {year}
  {1963})%
  \bibAnnoteFile{NoStop}{glauber_63}%
\bibitem{sudarshan_63}%
  \BibitemOpen
  \bibfield{author}{%
  \bibinfo {author} {\bibfnamefont{E.~C.~G.}\ \bibnamefont{Sudarshan}},\ }%
  \bibfield{journal}{%
  \bibinfo {journal} {Phys. Rev. Lett.}\ }%
  \textbf{\bibinfo {volume} {10}},\ \bibinfo {pages} {277} (\bibinfo {year}
  {1963})%
  \bibAnnoteFile{NoStop}{sudarshan_63}%
\bibitem{husimi_40}%
  \BibitemOpen
  \bibfield{author}{%
  \bibinfo {author} {\bibfnamefont{K.}~\bibnamefont{Husimi}},\ }%
  \bibfield{journal}{%
  \bibinfo {journal} {Proc. Phys. Math. Soc. Jpn}\ }%
  \textbf{\bibinfo {volume} {22}},\ \bibinfo {pages} {264} (\bibinfo {year}
  {1940})%
  \bibAnnoteFile{NoStop}{husimi_40}%
\bibitem{carter_drummond_87}%
  \BibitemOpen
  \bibfield{author}{%
  \bibinfo {author} {\bibfnamefont{S.~J.}\ \bibnamefont{Carter}}, \bibinfo
  {author} {\bibfnamefont{P.~D.}\ \bibnamefont{Drummond}}, \bibinfo {author}
  {\bibfnamefont{M.~D.}\ \bibnamefont{Reid}},\ and\ \bibinfo {author}
  {\bibfnamefont{R.~M.}\ \bibnamefont{Shelby}},\ }%
  \bibfield{journal}{%
  \bibinfo {journal} {Phys. Rev. Lett.}\ }%
  \textbf{\bibinfo {volume} {58}},\ \bibinfo {pages} {1841} (\bibinfo {year}
  {1987})%
  \bibAnnoteFile{NoStop}{carter_drummond_87}%
\bibitem{carter_drummond_91}%
  \BibitemOpen
  \bibfield{author}{%
  \bibinfo {author} {\bibfnamefont{S.~J.}\ \bibnamefont{Carter}}\ and\ \bibinfo
  {author} {\bibfnamefont{P.~D.}\ \bibnamefont{Drummond}},\ }%
  \bibfield{journal}{%
  \bibinfo {journal} {Phys. Rev. Lett.}\ }%
  \textbf{\bibinfo {volume} {67}},\ \bibinfo {pages} {3757} (\bibinfo {year}
  {1991})%
  \bibAnnoteFile{NoStop}{carter_drummond_91}%
\bibitem{drummond_hardman_93}%
  \BibitemOpen
  \bibfield{author}{%
  \bibinfo {author} {\bibfnamefont{P.~D.}\ \bibnamefont{Drummond}}\ and\
  \bibinfo {author} {\bibfnamefont{A.~D.}\ \bibnamefont{Hardman}},\ }%
  \bibfield{journal}{%
  \bibinfo {journal} {EuroPhys. Lett.}\ }%
  \textbf{\bibinfo {volume} {21}},\ \bibinfo {pages} {279} (\bibinfo {year}
  {1993})%
  \bibAnnoteFile{NoStop}{drummond_hardman_93}%
\bibitem{drummond_shelby_93}%
  \BibitemOpen
  \bibfield{author}{%
  \bibinfo {author} {\bibfnamefont{P.~D.}\ \bibnamefont{Drummond}}, \bibinfo
  {author} {\bibfnamefont{R.~M.}\ \bibnamefont{Shelby}}, \bibinfo {author}
  {\bibfnamefont{S.~R.}\ \bibnamefont{Friberg}},\ and\ \bibinfo {author}
  {\bibfnamefont{Y.}~\bibnamefont{Yamamoto}},\ }%
  \bibfield{journal}{%
  \bibinfo {journal} {Nature}\ }%
  \textbf{\bibinfo {volume} {365}},\ \bibinfo {pages} {307} (\bibinfo {year}
  {1993})%
  \bibAnnoteFile{NoStop}{drummond_shelby_93}%
\bibitem{Drummond2001a}%
  \BibitemOpen
  \bibfield{author}{%
  \bibinfo {author} {\bibfnamefont{P.~D.}\ \bibnamefont{Drummond}}\ and\
  \bibinfo {author} {\bibfnamefont{J.~F.}\ \bibnamefont{Corney}},\ }%
  \bibfield{journal}{%
  \bibinfo {journal} {J. Opt. Soc. Am. B}\ }%
  \textbf{\bibinfo {volume} {18}},\ \bibinfo {pages} {139} (\bibinfo {year}
  {2001})%
  \bibAnnoteFile{NoStop}{Drummond2001a}%
\bibitem{Rosenbluh1991a}%
  \BibitemOpen
  \bibfield{author}{%
  \bibinfo {author} {\bibfnamefont{M.}~\bibnamefont{Rosenbluh}}\ and\ \bibinfo
  {author} {\bibfnamefont{R.~M.}\ \bibnamefont{Shelby}},\ }%
  \bibfield{journal}{%
  \bibinfo {journal} {Phys. Rev. Lett.}\ }%
  \textbf{\bibinfo {volume} {66}},\ \bibinfo {pages} {153} (\bibinfo {year}
  {1991})%
  \bibAnnoteFile{NoStop}{Rosenbluh1991a}%
\bibitem{Heersink2005a}%
  \BibitemOpen
  \bibfield{author}{%
  \bibinfo {author} {\bibfnamefont{J.}~\bibnamefont{Heersink}}, \bibinfo
  {author} {\bibfnamefont{V.}~\bibnamefont{Josse}}, \bibinfo {author}
  {\bibfnamefont{G.}~\bibnamefont{Leuchs}},\ and\ \bibinfo {author}
  {\bibfnamefont{U.~L.}\ \bibnamefont{Andersen}},\ }%
  \bibfield{journal}{%
  \bibinfo {journal} {Opt. Lett.}\ }%
  \textbf{\bibinfo {volume} {30}},\ \bibinfo {pages} {1192} (\bibinfo {year}
  {2005})%
  \bibAnnoteFile{NoStop}{Heersink2005a}%
\bibitem{corney_drummond_06a}%
  \BibitemOpen
  \bibfield{author}{%
  \bibinfo {author} {\bibfnamefont{J.~F.}\ \bibnamefont{Corney}}, \bibinfo
  {author} {\bibfnamefont{P.~D.}\ \bibnamefont{Drummond}}, \bibinfo {author}
  {\bibfnamefont{J.}~\bibnamefont{Heersink}}, \bibinfo {author}
  {\bibfnamefont{V.}~\bibnamefont{Josse}}, \bibinfo {author}
  {\bibfnamefont{G.}~\bibnamefont{Leuchs}},\ and\ \bibinfo {author}
  {\bibfnamefont{U.~L.}\ \bibnamefont{Andersen}},\ }%
  \bibfield{journal}{%
  \bibinfo {journal} {Phys. Rev. Lett.}\ }%
  \textbf{\bibinfo {volume} {97}},\ \bibinfo {pages} {023606} (\bibinfo {year}
  {2006})%
  \bibAnnoteFile{NoStop}{corney_drummond_06a}%
\bibitem{steel_olsen_98}%
  \BibitemOpen
  \bibfield{author}{%
  \bibinfo {author} {\bibfnamefont{M.~J.}\ \bibnamefont{Steel}}, \bibinfo
  {author} {\bibfnamefont{M.~K.}\ \bibnamefont{Olsen}}, \bibinfo {author}
  {\bibfnamefont{L.~I.}\ \bibnamefont{Plimak}}, \bibinfo {author}
  {\bibfnamefont{P.~D.}\ \bibnamefont{Drummond}}, \bibinfo {author}
  {\bibfnamefont{S.~M.}\ \bibnamefont{Tan}}, \bibinfo {author}
  {\bibfnamefont{M.~J.}\ \bibnamefont{Collet}}, \bibinfo {author}
  {\bibfnamefont{D.~F.}\ \bibnamefont{Walls}},\ and\ \bibinfo {author}
  {\bibfnamefont{R.}~\bibnamefont{Graham}},\ }%
  \bibfield{journal}{%
  \bibinfo {journal} {Physical Review A}\ }%
  \textbf{\bibinfo {volume} {58}},\ \bibinfo {pages} {4824} (\bibinfo {year}
  {1998})%
  \bibAnnoteFile{NoStop}{steel_olsen_98}%
\bibitem{blakie_bradley_08}%
  \BibitemOpen
  \bibfield{author}{%
  \bibinfo {author} {\bibfnamefont{P.~B.}\ \bibnamefont{Blakie}}, \bibinfo
  {author} {\bibfnamefont{A.~S.}\ \bibnamefont{Bradley}}, \bibinfo {author}
  {\bibfnamefont{M.~J.}\ \bibnamefont{Davis}}, \bibinfo {author}
  {\bibfnamefont{R.~J.}\ \bibnamefont{Ballagh}},\ and\ \bibinfo {author}
  {\bibfnamefont{C.~W.}\ \bibnamefont{Gardiner}},\ }%
  \bibfield{journal}{%
  \bibinfo {journal} {Advances in Physics}\ }%
  \textbf{\bibinfo {volume} {57}},\ \bibinfo {pages} {363} (\bibinfo {year}
  {2008})%
  \bibAnnoteFile{NoStop}{blakie_bradley_08}%
\bibitem{Egorov2011}%
  \BibitemOpen
  \bibfield{author}{%
  \bibinfo {author} {\bibfnamefont{M.}~\bibnamefont{Egorov}}, \bibinfo {author}
  {\bibfnamefont{R.~P.}\ \bibnamefont{Anderson}}, \bibinfo {author}
  {\bibfnamefont{V.}~\bibnamefont{Ivannikov}}, \bibinfo {author}
  {\bibfnamefont{B.}~\bibnamefont{Opanchuk}}, \bibinfo {author}
  {\bibfnamefont{P.}~\bibnamefont{Drummond}}, \bibinfo {author}
  {\bibfnamefont{B.~V.}\ \bibnamefont{Hall}},\ and\ \bibinfo {author}
  {\bibfnamefont{A.~I.}\ \bibnamefont{Sidorov}},\ }%
  \bibfield{journal}{%
  \bibinfo {journal} {Phys. Rev. A}\ }%
  \textbf{\bibinfo {volume} {{84}}} (\bibinfo {month} {{AUG 22}}\ \bibinfo
  {year} {{2011}}),\ ISSN \bibinfo {issn} {{1050-2947}}%
  \bibAnnoteFile{NoStop}{Egorov2011}%
\bibitem{prx-131-2766}%
  \BibitemOpen
  \bibfield{author}{%
  \bibinfo {author} {\bibfnamefont{R.}~\bibnamefont{Glauber}},\ }%
  \bibfield{journal}{%
  \Doi{10.1103/PhysRev.131.2766}{\bibinfo {journal} {Physical Review}}\ }%
  \textbf{\bibinfo {volume} {131}},\ \bibinfo {pages} {2766} (\bibinfo {month}
  {Sep.}\ \bibinfo {year} {1963}),\ ISSN \bibinfo {issn} {0031-899X},\
  \url{http://link.aps.org/doi/10.1103/PhysRev.131.2766}%
  \bibAnnoteFile{NoStop}{prx-131-2766}%
\bibitem{pra-59-1538}%
  \BibitemOpen
  \bibfield{author}{%
  \bibinfo {author} {\bibfnamefont{K.}~\bibnamefont{Cahill}}\ and\ \bibinfo
  {author} {\bibfnamefont{R.}~\bibnamefont{Glauber}},\ }%
  \bibfield{journal}{%
  \Doi{10.1103/PhysRevA.59.1538}{\bibinfo {journal} {Physical Review A}}\ }%
  \textbf{\bibinfo {volume} {59}},\ \bibinfo {pages} {1538} (\bibinfo {month}
  {Feb.}\ \bibinfo {year} {1999}),\ ISSN \bibinfo {issn} {1050-2947},\
  \url{http://link.aps.org/doi/10.1103/PhysRevA.59.1538}%
  \bibAnnoteFile{NoStop}{pra-59-1538}%
\bibitem{pra-68-63822}%
  \BibitemOpen
  \bibfield{author}{%
  \bibinfo {author} {\bibfnamefont{J.~F.}\ \bibnamefont{Corney}}\ and\ \bibinfo
  {author} {\bibfnamefont{P.~D.}\ \bibnamefont{Drummond}},\ }%
  \bibfield{journal}{%
  \Doi{10.1103/PhysRevA.68.063822}{\bibinfo {journal} {Physical Review A}}\ }%
  \textbf{\bibinfo {volume} {68}},\ \bibinfo {pages} {23} (\bibinfo {month}
  {Aug.}\ \bibinfo {year} {2003}),\ ISSN \bibinfo {issn} {1050-2947},\
  \Eprint{http://arxiv.org/abs/0308064}{arXiv:0308064 [quant-ph]},\
  \url{http://link.aps.org/doi/10.1103/PhysRevA.68.063822
  http://arxiv.org/abs/quant-ph/0308064}%
  \bibAnnoteFile{NoStop}{pra-68-63822}%
\bibitem{Cahill1969}%
  \BibitemOpen
  \bibfield{author}{%
  \bibinfo {author} {\bibfnamefont{K.~E.}\ \bibnamefont{Cahill}}\ and\ \bibinfo
  {author} {\bibfnamefont{R.~J.}\ \bibnamefont{Glauber}},\ }%
  \bibfield{journal}{%
  \bibinfo {journal} {Phys. Rev.}\ }%
  \textbf{\bibinfo {volume} {177}},\ \bibinfo {pages} {1857} (\bibinfo {year}
  {1969})%
  \bibAnnoteFile{NoStop}{Cahill1969}%
\bibitem{moyal_49}%
  \BibitemOpen
  \bibfield{author}{%
  \bibinfo {author} {\bibfnamefont{J.~E.}\ \bibnamefont{Moyal}},\ }%
  \bibfield{journal}{%
  \bibinfo {journal} {Math.l Proc. Camb. Phil. Soc.}\ }%
  \textbf{\bibinfo {volume} {45}},\ \bibinfo {pages} {99} (\bibinfo {year}
  {1949})%
  \bibAnnoteFile{NoStop}{moyal_49}%
\bibitem{europhys-lett-21-279}%
  \BibitemOpen
  \bibfield{author}{%
  \bibinfo {author} {\bibfnamefont{P.~D.}\ \bibnamefont{Drummond}}\ and\
  \bibinfo {author} {\bibfnamefont{A.~D.}\ \bibnamefont{Hardman}},\ }%
  \bibfield{journal}{%
  \Doi{10.1209/0295-5075/21/3/005}{\bibinfo {journal} {Europhysics Letters}}\
  }%
  \textbf{\bibinfo {volume} {21}},\ \bibinfo {pages} {279} (\bibinfo {month}
  {jan}\ \bibinfo {year} {1993}),\ ISSN \bibinfo {issn} {0295-5075},\
  \url{http://stacks.iop.org/0295-5075/21/i=3/a=005?key=crossref.5ced39e5bf433%
e75a804fe992a5bc20e}%
  \bibAnnoteFile{NoStop}{europhys-lett-21-279}%
\bibitem{pra-58-4824}%
  \BibitemOpen
  \bibfield{author}{%
  \bibinfo {author} {\bibfnamefont{M.}~\bibnamefont{Steel}}, \bibinfo {author}
  {\bibfnamefont{M.~K.}\ \bibnamefont{Olsen}}, \bibinfo {author}
  {\bibfnamefont{L.~I.}\ \bibnamefont{Plimak}}, \bibinfo {author}
  {\bibfnamefont{P.~D.}\ \bibnamefont{Drummond}}, \bibinfo {author}
  {\bibfnamefont{S.}~\bibnamefont{Tan}}, \bibinfo {author}
  {\bibfnamefont{M.~J.}\ \bibnamefont{Collett}}, \bibinfo {author}
  {\bibfnamefont{D.}~\bibnamefont{Walls}},\ and\ \bibinfo {author}
  {\bibfnamefont{R.}~\bibnamefont{Graham}},\ }%
  \bibfield{journal}{%
  \Doi{10.1103/PhysRevA.58.4824}{\bibinfo {journal} {Physical Review A}}\ }%
  \textbf{\bibinfo {volume} {58}},\ \bibinfo {pages} {4824} (\bibinfo {month}
  {dec}\ \bibinfo {year} {1998}),\ ISSN \bibinfo {issn} {1050-2947},\
  \url{http://link.aps.org/doi/10.1103/PhysRevA.58.4824}%
  \bibAnnoteFile{NoStop}{pra-58-4824}%
\bibitem{jpb-35-3599}%
  \BibitemOpen
  \bibfield{author}{%
  \bibinfo {author} {\bibfnamefont{A.}~\bibnamefont{Sinatra}}, \bibinfo
  {author} {\bibfnamefont{C.}~\bibnamefont{Lobo}},\ and\ \bibinfo {author}
  {\bibfnamefont{Y.}~\bibnamefont{Castin}},\ }%
  \bibfield{journal}{%
  \Doi{10.1088/0953-4075/35/17/301}{\bibinfo {journal} {Journal of Physics B}}\
  }%
  \textbf{\bibinfo {volume} {35}},\ \bibinfo {pages} {3599} (\bibinfo {month}
  {sep}\ \bibinfo {year} {2002}),\ ISSN \bibinfo {issn} {0953-4075},\
  \url{http://stacks.iop.org/0953-4075/35/i=17/a=301?key=crossref.c7aed7eb9dc5%
ce0bb952f7815936c80e}%
  \bibAnnoteFile{NoStop}{jpb-35-3599}%
\bibitem{prl-89-140402}%
  \BibitemOpen
  \bibfield{author}{%
  \bibinfo {author} {\bibfnamefont{M.~W.}\ \bibnamefont{Jack}},\ }%
  \bibfield{journal}{%
  \Doi{10.1103/PhysRevLett.89.140402}{\bibinfo {journal} {Physical Review
  Letters}}\ }%
  \textbf{\bibinfo {volume} {89}},\ \bibinfo {pages} {140402} (\bibinfo {month}
  {sep}\ \bibinfo {year} {2002}),\ ISSN \bibinfo {issn} {0031-9007},\
  \url{http://link.aps.org/doi/10.1103/PhysRevLett.89.140402}%
  \bibAnnoteFile{NoStop}{prl-89-140402}%
\bibitem{nat-409-63}%
  \BibitemOpen
  \bibfield{author}{%
  \bibinfo {author} {\bibfnamefont{A.}~\bibnamefont{S{\o}rensen}}, \bibinfo
  {author} {\bibfnamefont{L.~M.}\ \bibnamefont{Duan}}, \bibinfo {author}
  {\bibfnamefont{J.~I.}\ \bibnamefont{Cirac}},\ and\ \bibinfo {author}
  {\bibfnamefont{P.}~\bibnamefont{Zoller}},\ }%
  \bibfield{journal}{%
  \Doi{10.1038/35051038}{\bibinfo {journal} {Nature}}\ }%
  \textbf{\bibinfo {volume} {409}},\ \bibinfo {pages} {63} (\bibinfo {month}
  {jan}\ \bibinfo {year} {2001}),\ ISSN \bibinfo {issn} {0028-0836},\
  \url{http://www.ncbi.nlm.nih.gov/pubmed/11343111}%
  \bibAnnoteFile{NoStop}{nat-409-63}%
\bibitem{pra-50-67}%
  \BibitemOpen
  \bibfield{author}{%
  \bibinfo {author} {\bibfnamefont{D.~J.}\ \bibnamefont{Wineland}}, \bibinfo
  {author} {\bibfnamefont{J.}~\bibnamefont{Bollinger}}, \bibinfo {author}
  {\bibfnamefont{W.}~\bibnamefont{Itano}},\ and\ \bibinfo {author}
  {\bibfnamefont{D.}~\bibnamefont{Heinzen}},\ }%
  \bibfield{journal}{%
  \Doi{10.1103/PhysRevA.50.67}{\bibinfo {journal} {Physical Review A}}\ }%
  \textbf{\bibinfo {volume} {50}},\ \bibinfo {pages} {67} (\bibinfo {month}
  {jul}\ \bibinfo {year} {1994}),\ ISSN \bibinfo {issn} {1050-2947},\
  \url{http://link.aps.org/doi/10.1103/PhysRevA.50.67}%
  \bibAnnoteFile{NoStop}{pra-50-67}%
\bibitem{pra-80-050701}%
  \BibitemOpen
  \bibfield{author}{%
  \bibinfo {author} {\bibfnamefont{A.}~\bibnamefont{Kaufman}}, \bibinfo
  {author} {\bibfnamefont{R.~P.}\ \bibnamefont{Anderson}}, \bibinfo {author}
  {\bibfnamefont{T.}~\bibnamefont{Hanna}}, \bibinfo {author}
  {\bibfnamefont{E.}~\bibnamefont{Tiesinga}}, \bibinfo {author}
  {\bibfnamefont{P.}~\bibnamefont{Julienne}},\ and\ \bibinfo {author}
  {\bibfnamefont{D.}~\bibnamefont{Hall}},\ }%
  \bibfield{journal}{%
  \Doi{10.1103/PhysRevA.80.050701}{\bibinfo {journal} {Physical Review A}}\ }%
  \textbf{\bibinfo {volume} {80}},\ \bibinfo {pages} {050701} (\bibinfo {month}
  {nov}\ \bibinfo {year} {2009}),\ ISSN \bibinfo {issn} {1050-2947},\
  \url{http://link.aps.org/doi/10.1103/PhysRevA.80.050701}%
  \bibAnnoteFile{NoStop}{pra-80-050701}%
\bibitem{drummond_gardiner_1980}%
  \BibitemOpen
  \bibfield{author}{%
  \bibinfo {author} {\bibfnamefont{P.~D.}\ \bibnamefont{Drummond}}\ and\
  \bibinfo {author} {\bibfnamefont{C.~W.}\ \bibnamefont{Gardiner}},\ }%
  \bibfield{journal}{%
  \bibinfo {journal} {J. Phys. A}\ }%
  \textbf{\bibinfo {volume} {13}},\ \bibinfo {pages} {2353} (\bibinfo {year}
  {1980})%
  \bibAnnoteFile{NoStop}{drummond_gardiner_1980}%
\bibitem{job-5-S281}%
  \BibitemOpen
  \bibfield{author}{%
  \bibinfo {author} {\bibfnamefont{P.~D.}\ \bibnamefont{Drummond}}\ and\
  \bibinfo {author} {\bibfnamefont{P.}~\bibnamefont{Deuar}},\ }%
  \bibfield{journal}{%
  \Doi{10.1088/1464-4266/5/3/359}{\bibinfo {journal} {Journal of Optics B:
  Quantum and Semiclassical Optics}}\ }%
  \textbf{\bibinfo {volume} {5}},\ \bibinfo {pages} {S281} (\bibinfo {month}
  {Jun.}\ \bibinfo {year} {2003}),\ ISSN \bibinfo {issn} {1464-4266},\
  \Eprint{http://arxiv.org/abs/0309514}{arXiv:0309514 [cond-mat]},\
  \url{http://arxiv.org/abs/cond-mat/0309514
  http://stacks.iop.org/1464-4266/5/i=3/a=359?key=crossref.9469b188d47f4d09fb9%
889dce3d7b81e}%
  \bibAnnoteFile{NoStop}{job-5-S281}%
\bibitem{dowling_drummond_05}%
  \BibitemOpen
  \bibfield{author}{%
  \bibinfo {author} {\bibfnamefont{M.~R.}\ \bibnamefont{Dowling}}, \bibinfo
  {author} {\bibfnamefont{P.~D.}\ \bibnamefont{Drummond}}, \bibinfo {author}
  {\bibfnamefont{M.~J.}\ \bibnamefont{Davis}},\ and\ \bibinfo {author}
  {\bibfnamefont{P.}~\bibnamefont{Deuar}},\ }%
  \bibfield{journal}{%
  \bibinfo {journal} {Phys. Rev. Lett.}\ }%
  \textbf{\bibinfo {volume} {94}},\ \bibinfo {pages} {130401} (\bibinfo {year}
  {2005})%
  \bibAnnoteFile{NoStop}{dowling_drummond_05}%
\bibitem{deuar_drummond_07}%
  \BibitemOpen
  \bibfield{author}{%
  \bibinfo {author} {\bibfnamefont{P.}~\bibnamefont{Deuar}}\ and\ \bibinfo
  {author} {\bibfnamefont{P.~D.}\ \bibnamefont{Drummond}},\ }%
  \bibfield{journal}{%
  \bibinfo {journal} {Phys. Rev. Lett.}\ }%
  \textbf{\bibinfo {volume} {98}},\ \bibinfo {pages} {120402} (\bibinfo {year}
  {2007})%
  \bibAnnoteFile{NoStop}{deuar_drummond_07}%
\bibitem{vogels_xu_02}%
  \BibitemOpen
  \bibfield{author}{%
  \bibinfo {author} {\bibfnamefont{J.~M.}\ \bibnamefont{Vogels}}, \bibinfo
  {author} {\bibfnamefont{K.}~\bibnamefont{Xu}},\ and\ \bibinfo {author}
  {\bibfnamefont{W.}~\bibnamefont{Ketterle}},\ }%
  \bibfield{journal}{%
  \bibinfo {journal} {Phys. Rev. Lett.}\ }%
  \textbf{\bibinfo {volume} {89}},\ \bibinfo {pages} {020401} (\bibinfo {year}
  {2002})%
  \bibAnnoteFile{NoStop}{vogels_xu_02}%
\bibitem{krachmalnicoff_jaskula_10}%
  \BibitemOpen
  \bibfield{author}{%
  \bibinfo {author} {\bibfnamefont{V.}~\bibnamefont{Krachmalnicoff}}, \bibinfo
  {author} {\bibfnamefont{J.-C.}\ \bibnamefont{Jaskula}}, \bibinfo {author}
  {\bibfnamefont{M.}~\bibnamefont{Bonneau}}, \bibinfo {author}
  {\bibfnamefont{V.}~\bibnamefont{Leung}}, \bibinfo {author}
  {\bibfnamefont{G.~B.}\ \bibnamefont{Partridge}}, \bibinfo {author}
  {\bibfnamefont{D.}~\bibnamefont{Boiron}}, \bibinfo {author}
  {\bibfnamefont{C.~I.}\ \bibnamefont{Westbrook}}, \bibinfo {author}
  {\bibfnamefont{P.}~\bibnamefont{Deuar}}, \bibinfo {author}
  {\bibfnamefont{P.}~\bibnamefont{Zin}}, \bibinfo {author}
  {\bibfnamefont{M.}~\bibnamefont{Trippenbach}},\ and\ \bibinfo {author}
  {\bibfnamefont{K.}~\bibnamefont{Kheruntsyan}},\ }%
  \bibfield{journal}{%
  \bibinfo {journal} {Phys. Rev. Lett.}\ }%
  \textbf{\bibinfo {volume} {104}},\ \bibinfo {pages} {150402} (\bibinfo {year}
  {2010})%
  \bibAnnoteFile{NoStop}{krachmalnicoff_jaskula_10}%
\bibitem{prl-98-120402}%
  \BibitemOpen
  \bibfield{author}{%
  \bibinfo {author} {\bibfnamefont{P.}~\bibnamefont{Deuar}}\ and\ \bibinfo
  {author} {\bibfnamefont{P.}~\bibnamefont{Drummond}},\ }%
  \bibfield{journal}{%
  \Doi{10.1103/PhysRevLett.98.120402}{\bibinfo {journal} {Physical Review
  Letters}}\ }%
  \textbf{\bibinfo {volume} {98}},\ \bibinfo {pages} {120402} (\bibinfo {month}
  {Mar.}\ \bibinfo {year} {2007}),\ ISSN \bibinfo {issn} {0031-9007},\
  \Eprint{http://arxiv.org/abs/0607831}{arXiv:0607831 [cond-mat]},\
  \url{http://arxiv.org/abs/cond-mat/0607831
  http://link.aps.org/doi/10.1103/PhysRevLett.98.120402}%
  \bibAnnoteFile{NoStop}{prl-98-120402}%
\bibitem{Corney:2003}%
  \BibitemOpen
  \bibfield{author}{%
  \bibinfo {author} {\bibfnamefont{J.~F.}\ \bibnamefont{Corney}}\ and\ \bibinfo
  {author} {\bibfnamefont{P.~D.}\ \bibnamefont{Drummond}},\ }%
  \bibfield{journal}{%
  \bibinfo {journal} {Phys. Rev. A}\ }%
  \textbf{\bibinfo {volume} {68}},\ \bibinfo {pages} {63822} (\bibinfo {year}
  {2003})%
  \bibAnnoteFile{NoStop}{Corney:2003}%
\bibitem{corney_drummond_06c}%
  \BibitemOpen
  \bibfield{author}{%
  \bibinfo {author} {\bibfnamefont{J.~F.}\ \bibnamefont{Corney}}\ and\ \bibinfo
  {author} {\bibfnamefont{P.~D.}\ \bibnamefont{Drummond}},\ }%
  \bibfield{journal}{%
  \bibinfo {journal} {Phys. Rev. B}\ }%
  \textbf{\bibinfo {volume} {73}},\ \bibinfo {pages} {125112} (\bibinfo {year}
  {2006})%
  \bibAnnoteFile{NoStop}{corney_drummond_06c}%
\bibitem{aimi_imada_07}%
  \BibitemOpen
  \bibfield{author}{%
  \bibinfo {author} {\bibfnamefont{T.}~\bibnamefont{Aimi}}\ and\ \bibinfo
  {author} {\bibfnamefont{M.}~\bibnamefont{Imada}},\ }%
  \bibfield{journal}{%
  \bibinfo {journal} {J. Phys. Soc. Jpn.}\ }%
  \textbf{\bibinfo {volume} {76}},\ \bibinfo {pages} {113708} (\bibinfo {year}
  {2007})%
  \bibAnnoteFile{NoStop}{aimi_imada_07}%
\bibitem{Renyi_1960}%
  \BibitemOpen
  \bibfield{author}{%
  \bibinfo {author} {\bibfnamefont{A.}~\bibnamefont{R\'enyi}},\ }%
  in\ \emph{\bibinfo {booktitle} {Proc. Fourth Berkeley Symp. Math. Stat. and
  Probability.}},\ Vol.~\bibinfo {volume} {I}\ (\bibinfo {publisher} {Berkeley,
  CA: University of California Press.},\ \bibinfo {year} {1961})\ pp.\ \bibinfo
  {pages} {547--561}%
  \bibAnnoteFile{NoStop}{Renyi_1960}%
\bibitem{ZaratePhysRevA2011}%
  \BibitemOpen
  \bibfield{author}{%
  \bibinfo {author} {\bibfnamefont{L.~E.~C.}\ \bibnamefont{Rosales-Z\'arate}}\
  and\ \bibinfo {author} {\bibfnamefont{P.~D.}\ \bibnamefont{Drummond}},\ }%
  \bibfield{journal}{%
  \Doi{10.1103/PhysRevA.84.042114}{\bibinfo {journal} {Phys. Rev. A}}\ }%
  \textbf{\bibinfo {volume} {84}},\ \bibinfo {pages} {042114} (\bibinfo {month}
  {Oct}\ \bibinfo {year} {2011}),\
  \url{http://link.aps.org/doi/10.1103/PhysRevA.84.042114}%
  \bibAnnoteFile{NoStop}{ZaratePhysRevA2011}%
\bibitem{Cahill1999a}%
  \BibitemOpen
  \bibfield{author}{%
  \bibinfo {author} {\bibfnamefont{K.~E.}\ \bibnamefont{Cahill}}\ and\ \bibinfo
  {author} {\bibfnamefont{R.~J.}\ \bibnamefont{Glauber}},\ }%
  \bibfield{journal}{%
  \bibinfo {journal} {Phys. Rev. A}\ }%
  \textbf{\bibinfo {volume} {59}},\ \bibinfo {pages} {1538} (\bibinfo {year}
  {1999})%
  \bibAnnoteFile{NoStop}{Cahill1999a}%
\bibitem{Frenkel-Wave}%
  \BibitemOpen
  \bibfield{author}{%
  \bibinfo {author} {\bibfnamefont{J.}~\bibnamefont{Frenkel}},\ }%
  \emph{\bibinfo {title} {{Wave Mechanics: Advanced General Theory}}},\
  \bibinfo {edition} {1st}\ ed.\ (\bibinfo {publisher} {Clarendon Press},\
  \bibinfo {address} {Oxford},\ \bibinfo {year} {1934})\
  \url{http://www.archive.org/details/wavemechanics030681mbp}%
  \bibAnnote{NoStop}{Frenkel-Wave}{The famous variational principle occurs on
  page 436.}%
\bibitem{Meyer-Multidimensional}%
  \BibitemOpen
  \emph{\bibinfo {title} {{Multidimensional Quantum Dynamics: MCTDH Theory and
  Applications}}},\ edited by\ \bibinfo {editor} {\bibfnamefont{H.-D.}\
  \bibnamefont{Meyer}}, \bibinfo {editor}
  {\bibfnamefont{F.}~\bibnamefont{Gatti}},\ and\ \bibinfo {editor}
  {\bibfnamefont{G.~A.}\ \bibnamefont{Worth}}\ (\bibinfo {publisher}
  {Wiley-VCH},\ \bibinfo {year} {2009})\ ISBN \bibinfo {isbn} {3527320180},\
  p.\ \bibinfo {pages} {442},\
  \url{http://www.amazon.com/Multidimensional-Quantum-Dynamics-Theory-Applicat%
ions/dp/3527320180}%
  \bibAnnoteFile{NoStop}{Meyer-Multidimensional}%
\bibitem{pra-77-33613}%
  \BibitemOpen
  \bibfield{author}{%
  \bibinfo {author} {\bibfnamefont{O.}~\bibnamefont{Alon}}, \bibinfo {author}
  {\bibfnamefont{A.}~\bibnamefont{Streltsov}},\ and\ \bibinfo {author}
  {\bibfnamefont{L.}~\bibnamefont{Cederbaum}},\ }%
  \bibfield{journal}{%
  \Doi{10.1103/PhysRevA.77.033613}{\bibinfo {journal} {Physical Review A}}\ }%
  \textbf{\bibinfo {volume} {77}},\ \bibinfo {pages} {33613} (\bibinfo {month}
  {Mar.}\ \bibinfo {year} {2008}),\ ISSN \bibinfo {issn} {1050-2947},\
  \url{http://link.aps.org/doi/10.1103/PhysRevA.77.033613}%
  \bibAnnoteFile{NoStop}{pra-77-33613}%
\bibitem{rmp-29-454}%
  \BibitemOpen
  \bibfield{author}{%
  \bibinfo {author} {\bibfnamefont{H.}~\bibnamefont{Everett}},\ }%
  \bibfield{journal}{%
  \Doi{10.1103/RevModPhys.29.454}{\bibinfo {journal} {Reviews of Modern
  Physics}}\ }%
  \textbf{\bibinfo {volume} {29}},\ \bibinfo {pages} {454} (\bibinfo {month}
  {Jul.}\ \bibinfo {year} {1957}),\ ISSN \bibinfo {issn} {0034-6861},\
  \url{http://link.aps.org/doi/10.1103/RevModPhys.29.454}%
  \bibAnnoteFile{NoStop}{rmp-29-454}%
\bibitem{jcp-128-044116}%
  \BibitemOpen
  \bibfield{author}{%
  \bibinfo {author} {\bibfnamefont{T.}~\bibnamefont{Fabcic}}, \bibinfo {author}
  {\bibfnamefont{J.}~\bibnamefont{Main}},\ and\ \bibinfo {author}
  {\bibfnamefont{G.}~\bibnamefont{Wunner}},\ }%
  \bibfield{journal}{%
  \Doi{10.1063/1.2821751}{\bibinfo {journal} {The Journal of chemical
  physics}}\ }%
  \textbf{\bibinfo {volume} {128}},\ \bibinfo {pages} {044116} (\bibinfo
  {month} {Jan.}\ \bibinfo {year} {2008}),\ ISSN \bibinfo {issn} {0021-9606},\
  \url{http://www.ncbi.nlm.nih.gov/pubmed/18247939}%
  \bibAnnoteFile{NoStop}{jcp-128-044116}%
\bibitem{Tikhonov-Nonlinear}%
  \BibitemOpen
  \bibfield{author}{%
  \bibinfo {author} {\bibfnamefont{A.~N.}\ \bibnamefont{Tikhonov}}, \bibinfo
  {author} {\bibfnamefont{A.~S.}\ \bibnamefont{Leonov}},\ and\ \bibinfo
  {author} {\bibfnamefont{A.~G.}\ \bibnamefont{Yagola}},\ }%
  \emph{\bibinfo {title} {{Nonlinear Ill-posed Problems}}}\ (\bibinfo
  {publisher} {Chapman \& Hall},\ \bibinfo {address} {London},\ \bibinfo {year}
  {1990})%
  \bibAnnoteFile{NoStop}{Tikhonov-Nonlinear}%
\bibitem{pra-33-674}%
  \BibitemOpen
  \bibfield{author}{%
  \bibinfo {author} {\bibfnamefont{G.}~\bibnamefont{Milburn}},\ }%
  \bibfield{journal}{%
  \Doi{10.1103/PhysRevA.33.674}{\bibinfo {journal} {Physical Review A}}\ }%
  \textbf{\bibinfo {volume} {33}},\ \bibinfo {pages} {674} (\bibinfo {month}
  {Jan.}\ \bibinfo {year} {1986}),\ ISSN \bibinfo {issn} {0556-2791},\
  \url{http://link.aps.org/doi/10.1103/PhysRevA.33.674}%
  \bibAnnoteFile{NoStop}{pra-33-674}%
\bibitem{Greiner_02}%
  \BibitemOpen
  \bibfield{author}{%
  \bibinfo {author} {\bibfnamefont{M.}~\bibnamefont{Greiner}}, \bibinfo
  {author} {\bibfnamefont{O.}~\bibnamefont{Mandel}}, \bibinfo {author}
  {\bibfnamefont{T.~W.}\ \bibnamefont{Hansch}},\ and\ \bibinfo {author}
  {\bibfnamefont{I.}~\bibnamefont{Bloch}},\ }%
  \bibfield{journal}{%
  \bibinfo {journal} {Nature}\ }%
  \textbf{\bibinfo {volume} {419}},\ \bibinfo {pages} {51 } (\bibinfo {year}
  {2002})%
  \bibAnnoteFile{NoStop}{Greiner_02}%
\bibitem{Deuar:2002}%
  \BibitemOpen
  \bibfield{author}{%
  \bibinfo {author} {\bibfnamefont{P.}~\bibnamefont{Deuar}}\ and\ \bibinfo
  {author} {\bibfnamefont{P.~D.}\ \bibnamefont{Drummond}},\ }%
  \bibfield{journal}{%
  \bibinfo {journal} {Phys. Rev. A}\ }%
  \textbf{\bibinfo {volume} {66}},\ \bibinfo {pages} {33812} (\bibinfo {year}
  {2002})%
  \bibAnnoteFile{NoStop}{Deuar:2002}%
\bibitem{deuar_drummond_06}%
  \BibitemOpen
  \bibfield{author}{%
  \bibinfo {author} {\bibfnamefont{P.}~\bibnamefont{Deuar}}\ and\ \bibinfo
  {author} {\bibfnamefont{P.~D.}\ \bibnamefont{Drummond}},\ }%
  \bibfield{journal}{%
  \bibinfo {journal} {Journal of Physics A}\ }%
  \textbf{\bibinfo {volume} {39}},\ \bibinfo {pages} {2723 } (\bibinfo {year}
  {2006})%
  \bibAnnoteFile{NoStop}{deuar_drummond_06}%
\end{thebibliography}%

\end{document}